\documentclass[preprint]{aastex}
\usepackage{lineno}

\shorttitle{First combined search for neutrino point-sources in the Southern Hemisphere with the ANTARES and IceCube neutrino telescopes}
\shortauthors{ANTARES and IceCube Collaborations}

\begin{document}

\title{First combined search for neutrino point-sources in the Southern Hemisphere with the ANTARES and IceCube neutrino telescopes}

\author{ANTARES Collaboration:
S.~Adri\'an-Mart\'inez\altaffilmark{1},
A.~Albert\altaffilmark{2},
M.~Andr\'e\altaffilmark{3},
G.~Anton\altaffilmark{4},
M.~Ardid\altaffilmark{1},
J.-J.~Aubert\altaffilmark{5},
B.~Baret\altaffilmark{6},
J.~Barrios-Mart\'{\i}\altaffilmark{7},
S.~Basa\altaffilmark{8},
V.~Bertin\altaffilmark{5}
% ...
}

\altaffiltext{1}{{Institut d'Investigaci\'o per a la Gesti\'o Integrada de les Zones Costaneres (IGIC) - Universitat Polit\`ecnica de Val\`encia. C/  Paranimf 1 , 46730 Gandia, Spain.}}
\altaffiltext{2}{{GRPHE - Universit\'e de Haute Alsace - Institut universitaire de technologie de Colmar, 34 rue du Grillenbreit BP 50568 - 68008 Colmar, France}}
\altaffiltext{3}{{Technical University of Catalonia, Laboratory of Applied Bioacoustics, Rambla Exposici\'o,08800 Vilanova i la Geltr\'u,Barcelona, Spain}}
\altaffiltext{4}{{Friedrich-Alexander-Universit\"at Erlangen-N\"urnberg, Erlangen Centre for Astroparticle Physics, Erwin-Rommel-Str. 1, 91058 Erlangen, Germany}}
\altaffiltext{5}{{Aix-Marseille Universit\'e, CNRS/IN2P3, CPPM UMR 7346, 13288 Marseille, France}}
\altaffiltext{6}{{APC, Universit\'e Paris Diderot, CNRS/IN2P3, CEA/IRFU, Observatoire de Paris, Sorbonne Paris Cit\'e, 75205 Paris, France}}
\altaffiltext{7}{{IFIC - Instituto de F\'isica Corpuscular, Edificios Investigaci\'on de Paterna, CSIC - Universitat de Val\`encia, Apdo. de Correos 22085, 46071 Valencia, Spain}}
\altaffiltext{8}{{LAM - Laboratoire d'Astrophysique de Marseille, P\^ole de l'\'Etoile Site de Ch\^ateau-Gombert, rue Fr\'ed\'eric Joliot-Curie 38,  13388 Marseille Cedex 13, France}}
% ...

\begin{abstract}

We present the results of searches for point-like sources of neutrinos based on the first combined analysis of data from both the ANTARES and IceCube neutrino telescopes.
 The combination of both detectors which differ in size and location forms a window in the Southern sky where the sensitivity to point sources improves by up to a factor of two compared to individual analyses.  Using data recorded by ANTARES from 2007 to 2012, and by IceCube from 2008 to 2011, we search for sources of neutrino emission both across the Southern sky and from a pre-selected list of candidate objects.  No significant excess over background has been found in these searches, and flux upper limits for the candidate sources are presented for $E^{-2.5}$ and $E^{-2}$ power-law spectra with different energy cut-offs.

\end{abstract}

\keywords{neutrino telescopes, neutrino astronomy, ANTARES, IceCube}

\section{Introduction}

Neutrinos offer unique insight into the Universe due to the fact that they interact only weakly and via gravity. Unlike charged particles, they can travel straight from the source to the Earth without being deflected by magnetic fields or being absorbed. Neutrinos are expected to originate from the same locations where the acceleration of cosmic rays take place \citep{Origin6, Origin4, Origin3, Origin2, Origin5, Origin}. A large variety of classes of astrophysical objects are predicted to be sources of high energy neutrinos, where galactic candidates include microquasars \citep{MicroQ1, MicroQ2, MicroQ3, MicroQ4}, supernova remnants \citep{SNRs1, SNRs2, SNRs3, SNRs4, SNRs5, SNRs6, SNRs7}, or various objects close to the Galactic Center \citep{SgrA, GC}. Extragalactic sources comprise active galactic nuclei \citep{AGN1, AGN2, AGN3, AGN4, AGN5, AGN6, AGN7, AGN8, AGN9} and gamma ray bursts \citep{GRB1, GRB2, GRB3, GRB4, GRB5, GRB6, GRB7}.

The low neutrino cross section also implies that their detection is challenging. After the pioneering efforts by the Baikal \citep{baikal} and AMANDA \citep{amanda} collaborations, the field is presently led by the IceCube \citep{icecube} and ANTARES \citep{AntDetect} experiments. IceCube, which is placed in the deep Antartic ice, is the first detector to reach the cubic-kilometer size predicted to be necessary to detect cosmic neutrino fluxes according to the Waxman-Bahcall flux \citep{Waxman}. Recently, IceCube has reported the crucial discovery of a flux of neutrinos  up to $\sim$\,PeV energies which cannot be explained by the background of atmospheric muons and neutrinos only \citep{
IceCube_HESE_2yr, IceCube_HESE_3yr}. The specific origin of these events is currently unknown. Some authors propose that at least part of the flux may have a galactic origin \citep{GAL-IC, GAL-IC2, GAL-IC3, GAL-IC4, GAL-IC5, GAL-IC6, GAL-IC7}, whereas others have focused on the extragalactic component \citep{EXTRAGAL1, EXTRAGAL2, EXTRAGAL3, EXTRAGAL4}. Meanwhile the ANTARES experiment has proven the feasibility of the Cherenkov telescope technique in sea water \citep{ANTARES-OSCI, ANTARES-MUAT}.
While its instrumented volume is significantly smaller than that of IceCube, its geographical location provides a  view of the Southern sky with significantly reduced background for neutrino energies below 100 TeV, and hence better sensitivity to many predicted Galactic sources of neutrinos in this part of the sky.  The complementarity of the detectors with respect to Southern sky sources, due to their different geographical location, size and atmospheric muon background, allows for a gain in sensitivity by combining the analyses of data from both experiments in a joint search for point-like sources.
The level of improvement depends on the details of the assumed astrophysical flux, in particular on its energy spectrum and the existence of a possible high-energy cut-off. The energy spectra are not yet known and predictions vary widely depending on the source model.

In this paper, a combined analysis using the point-source data samples of IceCube from 2008-2011 and of ANTARES from 2007-2012 is presented. This paper is structured as follows: in Section~\ref{ANTARESdet}, the IceCube and ANTARES detectors are introduced. In Section~\ref{sample} the samples from each experiment are described, while in Section~\ref{method} the search method is explained. Finally, the results are presented in Section~\ref{results} and the conclusions are discussed in Section~\ref{conclusions}. 


\section{The IceCube and ANTARES neutrino telescopes}\label{ANTARESdet}

IceCube is a cubic-kilometer neutrino telescope located at the geographic South Pole.  It consists of a total of 5160 digital optical modules (DOMs) deployed in the Antarctic ice at depths from 1450\,m to 2450\,m below the surface \citep{IceCube_DOM}. Each DOM consists of a pressure-resistant sphere that houses electronics, calibration LEDs, and a 10'' PMT facing downward. The DOMs are configured in a hexagonal array of 86 vertical cables descending from the surface, called ``strings'', with 60 DOMs per string.  The average horizontal distance is 125\,m between strings, and average vertical spacing is 17\,m between DOMs on a string. A sub-array of eight strings (Deep Core) is also present in the core of the detector \citep{DeepCore}. These strings have a smaller separation in order to improve the sensitivity for lower energies.  Construction of the detector began in 2005 and was completed six years later. The analysis presented in this paper is based on data from three years of the partially completed detector, when 40, 59, and 79 strings were deployed.  Future joint analyses are envisioned that will be based on data from the full 86-string detector, including recent data samples that use outer detector modules as vetoes to achieve sensitivity to lower energy neutrinos.

ANTARES is the first neutrino telescope which operates in the sea \citep{AntDetect}. It was completed in 2008, with the first lines operating from 2006. It is located in the Mediterranean Sea at a depth of 2475 m, at coordinates (42$^\circ$ 48' N, 6$^\circ$ 10' E), 40 km South of Toulon (France). It consists of an array of 885 Optical Modules (OMs) distributed along 12 lines of 350 m height and an inter-line separation of 60 to 75 m. An OM consists of a 10'' photomultiplier tube (PMT) contained inside a 17'' glass sphere. The OMs are grouped into triplets and face downward at an angle of 45$^\circ$ in order to optimize the detection of up-going muon-neutrinos. There are 25 triplets (storeys) on each line, with a distance of about 15 m between storeys. Lines are kept vertical with a buoy at their top. 

One of the main focuses of the ANTARES and IceCube neutrino telescopes is the observation of cosmic point-like sources of neutrino emission. At present, corresponding searches are mainly focused on the detection of muon neutrinos, which can be reconstructed with sub-degree angular resolution. Muon neutrinos are indirectly detected through the muon produced in their charged current interaction (CC) with a nucleus (N) inside or near to the detector volume:

\begin{equation}
\nu_\mu + N \rightarrow \mu^- + X 
\end{equation}

In this reaction, a muon and a hadronic shower, X, are produced.\footnote{In this work the charge conjugate particles and reactions will be implicitly included, i.e. in this case the reaction $\bar{\nu}_\mu + N \rightarrow \mu^+ + X$ is also assumed.} The ultra-relativistic muon can travel long distances (up to several kilometers) and, when crossing a suitable medium such as ice or water, induce Cherenkov radiation that can be detected by the photomultipliers (PMTs) of neutrino telescopes. The corresponding charge and time information of the detected photons is used to reconstruct the direction of the muon, which is almost collinear with the original neutrino for energies above the TeV range. The main backgrounds for cosmic neutrino searches are atmospheric muons and neutrinos produced in the decay of the secondary particles created in the interactions of cosmic rays with the nuclei of the atmosphere.

\section{Neutrino Data Samples}\label{sample}

The data sample employed for this analysis corresponds to all events from the Southern sky which were included in the three-year IceCube point-source analysis \citep{PS-IceCube-79} combined with the events in the latest ANTARES point-source analysis \citep{PS-ANTARES}.  The ANTARES sample contains data recorded from  Jan 29, 2007 to Dec 31, 2012; for IceCube the data was recorded from Apr 5, 2008 to May 13, 2011 with the partially completed detector, and without the use of the Deep Core strings.

\begin{figure}[!ht]
	\centering
	\plottwo{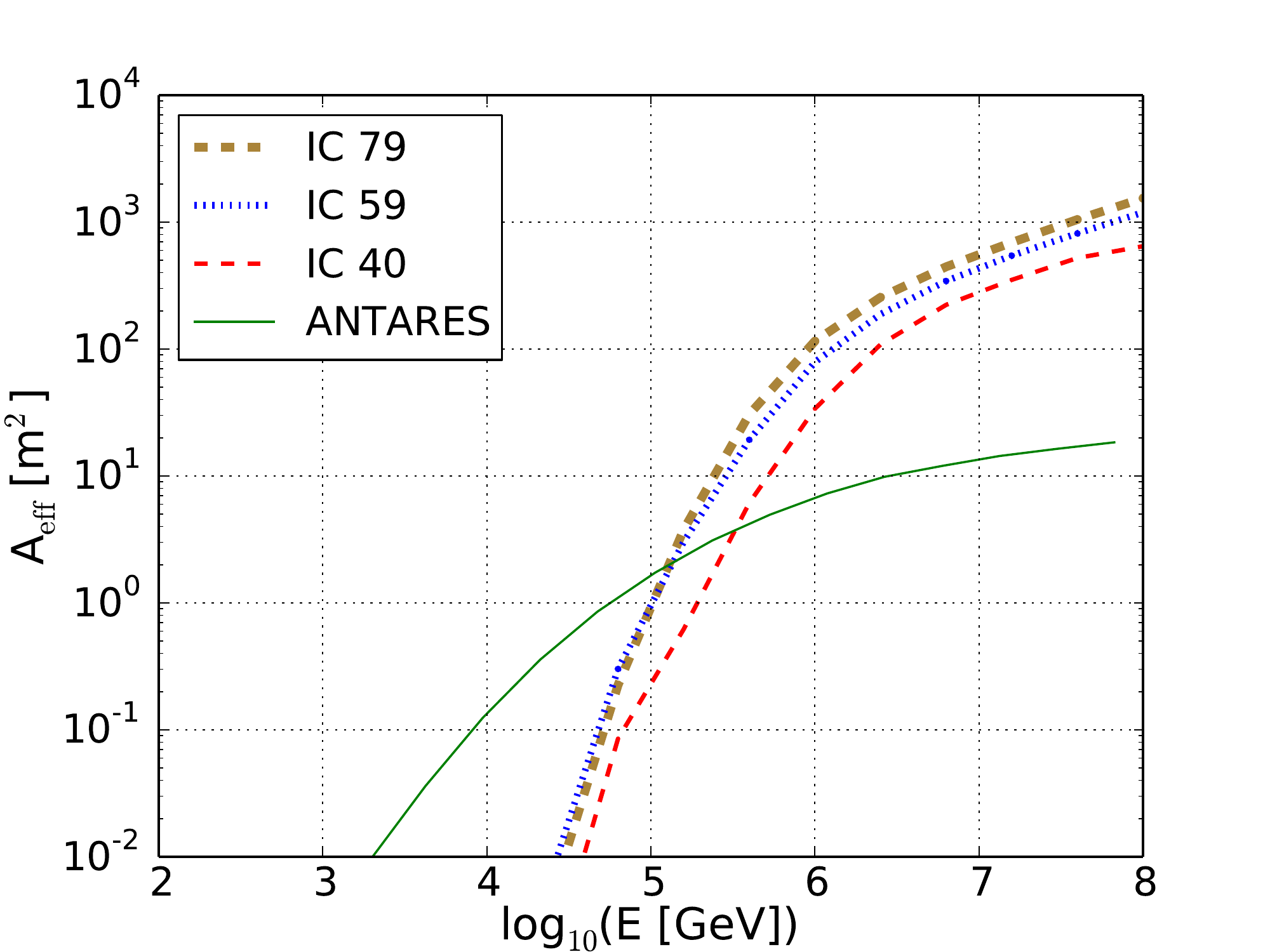}{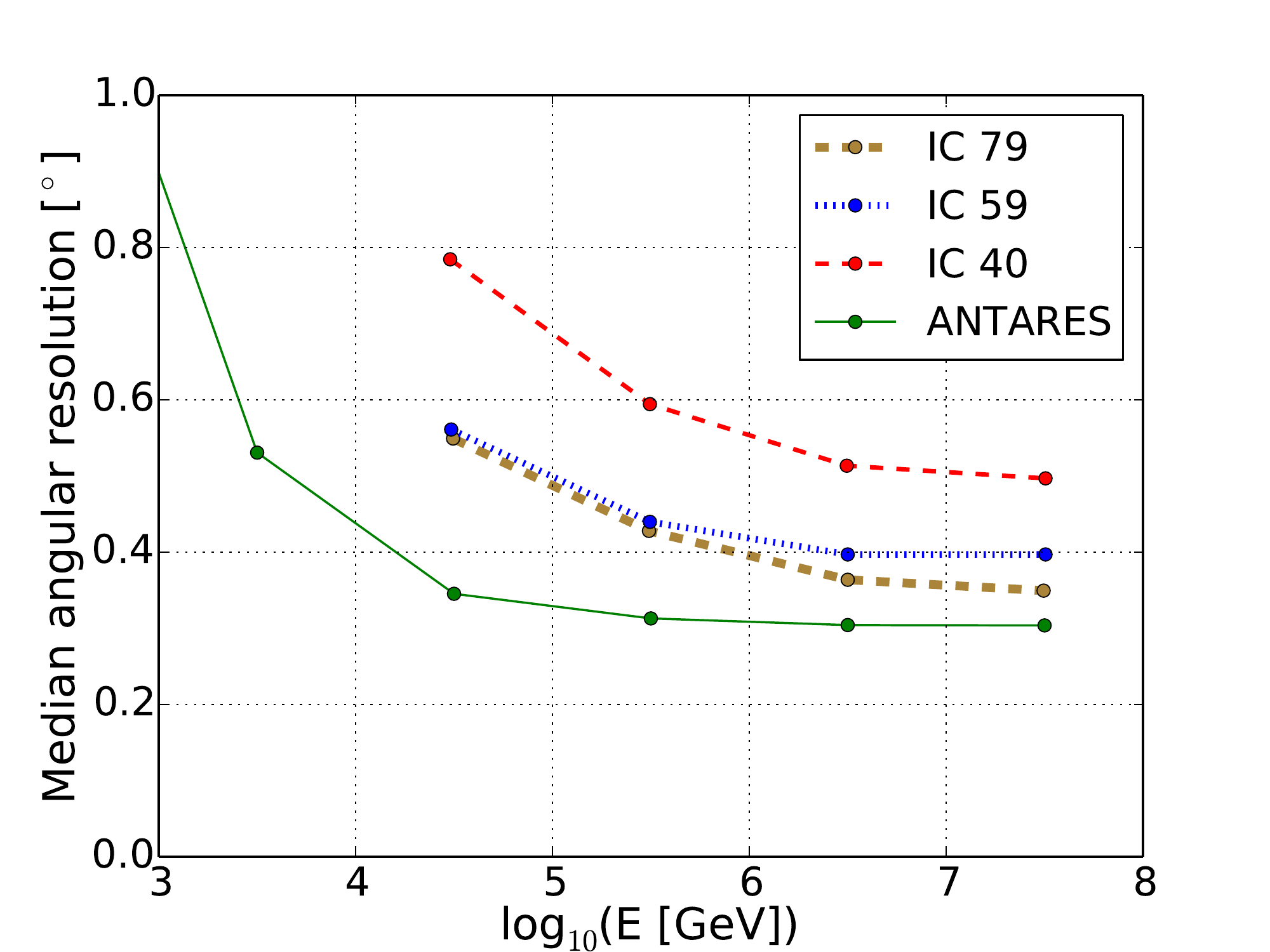}
	\caption{Muon neutrino effective area for a point source at a declination $\delta$ = --30$^\circ$ (left) and median angular resolution (right) for the samples used in this analysis after the final set of cuts. The median angular resolution is defined as the median of the difference between the true neutrino direction and the reconstructed muon direction.} 
	\label{Aeff}
\end{figure}

\begin{figure}[!ht]
	\centering
	\includegraphics[width=.8\textwidth]{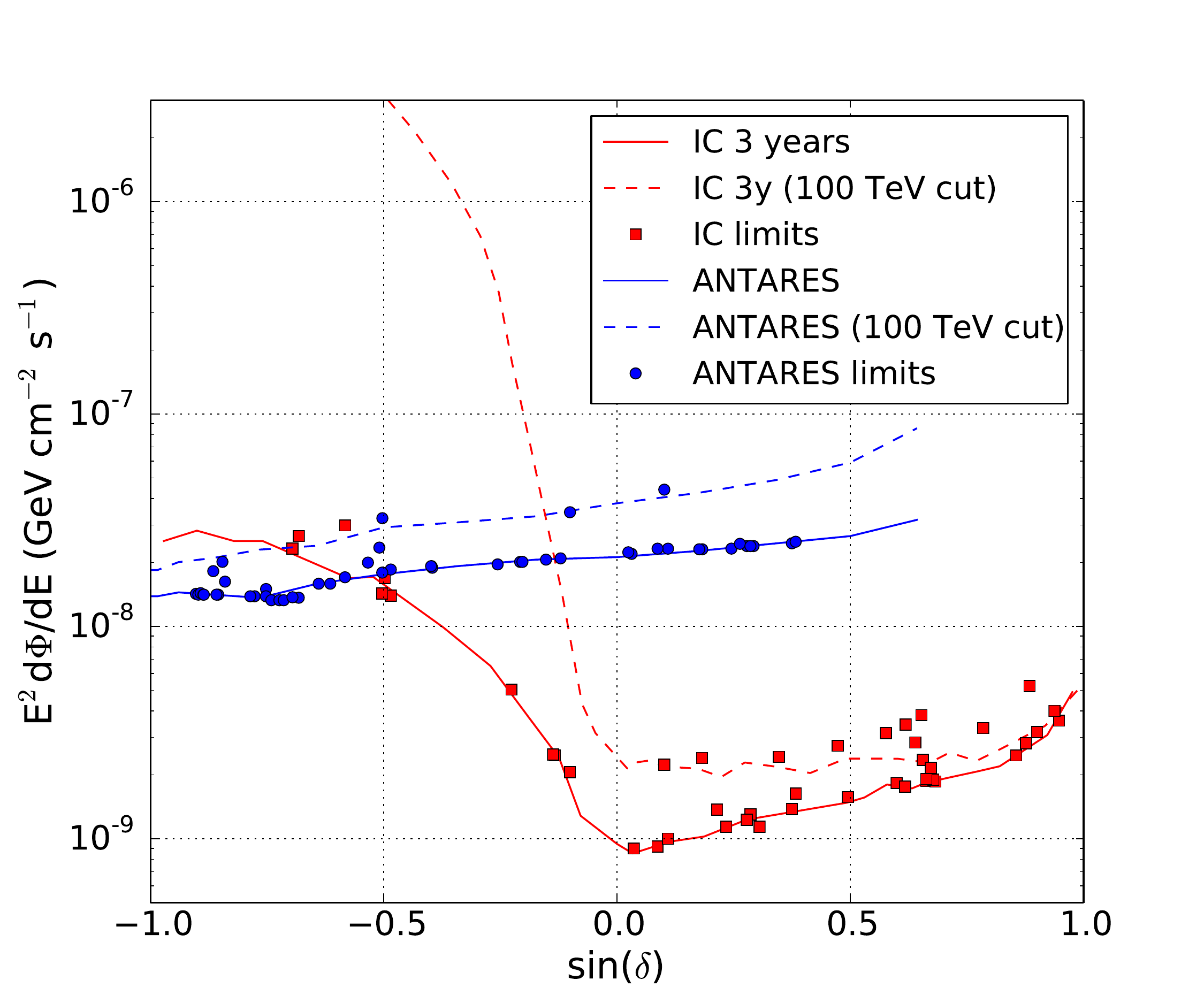}
	\caption{90\% CL limits for selected sources (squares and dots) and sensitivities using the Neyman method as a function of the declination (lines) reported in the ANTARES 2007-2012 (blue) \citep{PS-ANTARES} and the IceCube 3 years (red) \citep{PS-IceCube-79} point source analyses. An unbroken $E^{-2}$ power-law source spectrum is assumed for the limits and lower sensitivity curves (solid lines). Dashed lines indicate the sensitivity for an $E^{-2}$ spectrum with neutrino energies of $E_\nu \leq$ 100 TeV using the Neyman method. }
	\label{sensComp}
\end{figure}

Detector performance differs not only between ANTARES and IceCube, but also between the three IceCube configurations as the detector grew, from 40, to 59, and then to 79 strings. The effective area is defined as the equivalent surface with a perfect efficiency which detects the same number of events as the detector.
For a source position of $\delta$ = --30$^\circ$, the effective area for each IceCube configuration and for ANTARES is shown in Figure \ref{Aeff}-left.
 Due to its larger size, the effective area for the IceCube samples is larger for neutrino energies above $\sim$ 100 TeV. However, to view sources in the Southern Sky, IceCube must contend with the down-going background of atmospheric muons, which becomes overwhelming at lower energies. To minimize these, the IceCube point-source analysis introduced a declination-dependent energy cut which strongly suppresses low energy events in the final data sample. ANTARES, which can use the Earth as a filter against atmospheric muons in the Southern sky, thus maintains a larger effective area in this energy and declination range. 

A comparison of the median angular resolution of each sample can be seen in Figure \ref{Aeff}-right. The better resolution of the ANTARES sample is due to the longer photon scattering length in water compared to ice. The sensitivities reported by both experiments for the whole sky using the Neyman method \citep{Neyman} are shown in Figure \ref{sensComp}.
 
Different selection criteria are applied to each sample. A summary of these selections, which in all cases were developed with a data blinding policy and were optimised to minimise the neutrino flux needed for a 5$\sigma$ discovery in 50\% of the experiments, is given below.

\subsection{ANTARES}

The ANTARES data sample used for this analysis corresponds to the events coming from the Southern Sky used in the last published point source analysis \citep{PS-ANTARES}. The parameters which are used to optimise this sample are the quality of the track fit, $\Lambda$, the angular error estimate, $\sigma$ (also denoted as $\beta$ in most ANTARES publications), and the zenith angle, $\theta$. These three parameters are given by the track reconstruction of neutrino events, which uses a maximum likelihood (ML) method \citep{AartThesis, OtherAart}. The algorithm is based on a multi-step procedure to fit the direction of the reconstructed muon by maximising the $\Lambda$ parameter. The angular error estimate, $\sigma$, is obtained from the uncertainty on the zenith and azimuth angles extracted from the covariance matrix.

The selection yields a total of 5516 events for the whole sky, with 4136 of these events in the Southern Hemisphere. The estimated contamination of mis-reconstructed atmospheric muons is 10\%.

\subsection{IceCube}

\begin{deluxetable}{ccccc}
\tablewidth{0pt}
\tablecaption{Event samples for the different IceCube detector configurations, labelled by the number of strings deployed.  Only Southern-sky events (numbers indicated by last column) have been selected for the present analysis. }
\tablehead{
\colhead{Sample}  & \colhead{Start date}  & \colhead{End date} & \colhead{Livetime [days]}  & \colhead{\# events}
}
\startdata
IC-40 & 2008 Apr 5 &  2009 May 20 &  376 & 22 779\\
IC-59 & 2009 May 20 & 2010 May 31 & 348 & 64 230\\
IC-79 & 2010 May 31 & 2011 May 13 & 316 & 59 009\\
\enddata
\label{tab:icecube_events}
\end{deluxetable}

The IceCube data samples used for this analysis are based on the event selection optimized for point source searches with the data recorded using the 40, 59, and 79-string detector configurations, summarized in Table~\ref{tab:icecube_events}. Only events from the Southern sky are selected here for the joint analysis. In contrast to the ANTARES selection above, IceCube's Southern sky events are predominantly atmospheric muons rather than atmospheric neutrinos, because the Earth cannot be used as a neutrino filter for directions above the detector. 

The total number of down-going events in IceCube is $\sim 10^{10}$ per year.  The down-going events that were selected as part of the above analyses and which are used here comprise only well-reconstructed muon tracks at very high energies, where it becomes possible to detect a neutrino source with a hard $E^{-2}$ energy spectrum beyond the more steeply falling atmospheric muon background, and from clustering of events in a single region of the sky. For the 40 string configuration, a set of cuts on the reduced log-likelihood of the track reconstruction, the angular uncertainty, $\sigma$, and the muon energy proxy is performed for events coming from the Southern Sky \citep{PS-IceCube-40}. For the 59 string configuration, the vetoing capability of IceTop is added to reduce the background of atmospheric muons \citep{PS-IceCube-79}. For the 79 string configuration, the event selection is performed based on boosted decision trees using 17 observables, and includes the use of the IceTop veto.

The total number of Southern sky events selected from the three year sample is 146\,018 events.

\subsection{Relative fraction of source events for different source assumptions}\label{Sec:Relative}

The relative fraction of expected source events from each sample needs to be calculated in order to estimate its respective weight in the likelihood which will be used to search for an excess of events from a particular direction (see Section \ref{method}). This fraction is defined as the ratio of the expected number of signal events for the given sample to that for all samples,

\begin{equation}
C^j (\delta, {d\Phi}/{dE_\nu} ) = \frac{N^j(\delta, {d\Phi}/{dE_\nu})}{\sum_{i} N^i(\delta, {d\Phi}/{dE_\nu})} ,
\end{equation}

where the total number of expected events for the $j$-th sample, $N^j$, with a given source declination, $\delta$, and a given source spectrum, $\frac{d\Phi}{dE_\nu}$, can be calculated as

\begin{equation}\label{eq:acc}
N^j \left(\delta, \frac{d\Phi}{dE_\nu} \right) = \int dt \int dE_\nu A^{j}_{\rm eff}(E_\nu,\delta) \frac{d\Phi}{dE_\nu} \; .
\end{equation}

The time integration extends over the live time of each sample and $A^{j}_{\mathrm{eff}}(E_\nu, \delta)$ indicates the effective area of the corresponding detector layout $j$ as a function of the neutrino energy, $E_\nu$, and the declination of the source, $\delta$. 

\begin{figure}[!h]
	\centering
	\includegraphics[width=.7\textwidth]{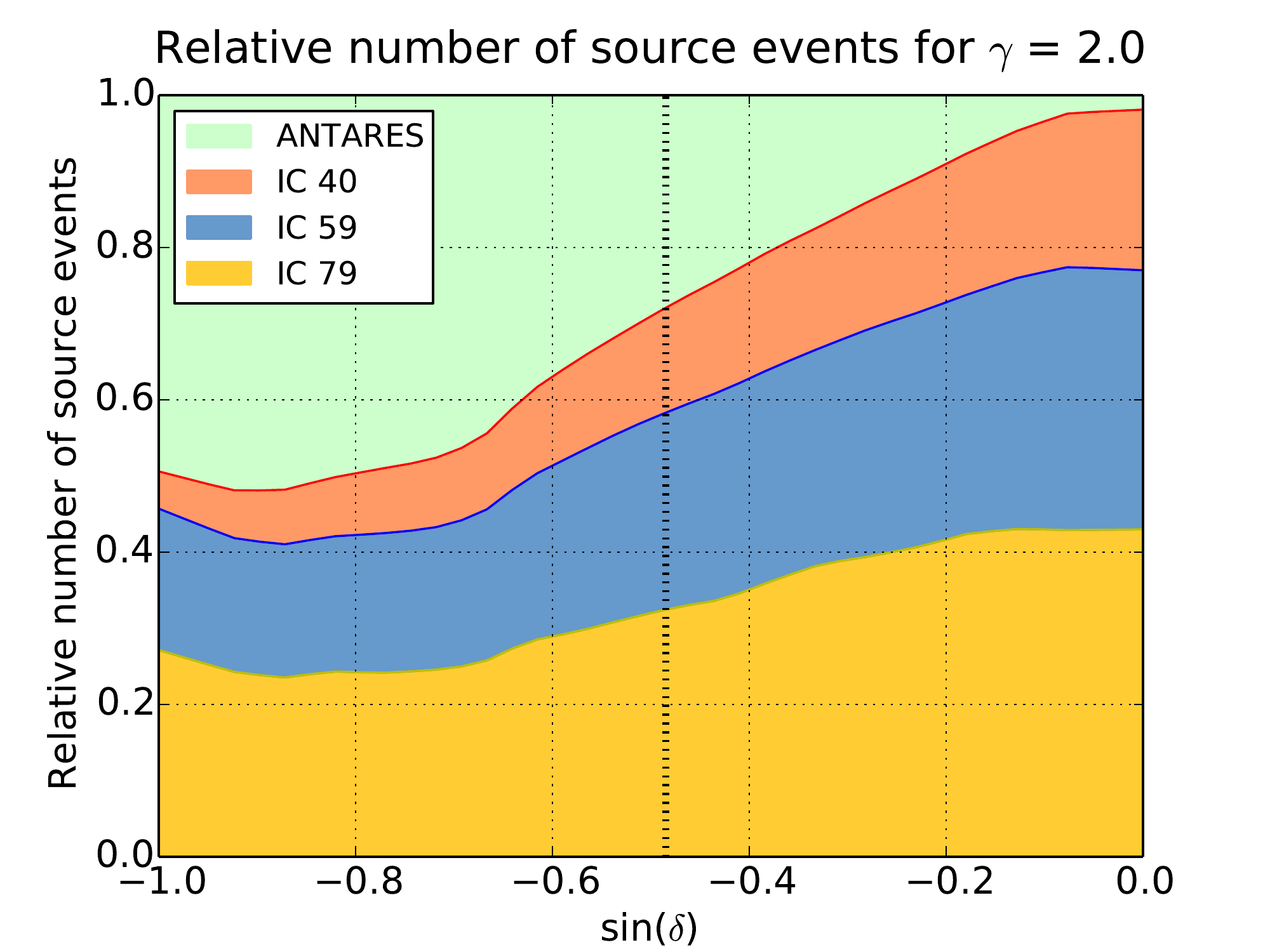}
	\caption{Relative fraction of signal events for each sample as a function of the source declination for the case of an $E^{-2}$ energy spectrum. The orange, blue, and yellow shaded areas correspond to the IceCube 40, 59 and 79-string data samples, respectively, and the green shaded area indicates the ANTARES 2007-2012 sample. The vertical dashed line marks the declination of the Galactic Center. }
	\label{RelContX20}
\end{figure}

Since each detector layout has a different response depending on the neutrino energy and declination, this relative fraction of source events needs to be calculated for different source spectra and source declinations. Figure \ref{RelContX20} shows the relative fraction of signal events for an unbroken $E^{-2}$ spectrum, which corresponds to vanilla first order Fermi acceleration \citep{Fermi1, Fermi2}. In this case, there is a significant contribution from all samples over most of the Southern Sky, with the ANTARES contribution being more significant for declinations closer to $\delta$ = --90$^\circ$, and IceCube for declinations closer to 0$^\circ$. The reason for this variability is mostly due to the declination-dependent energy cut applied in the IceCube samples to reduce the background of atmospheric muons. 

\begin{figure}[!h]
	\centering
	\plottwo{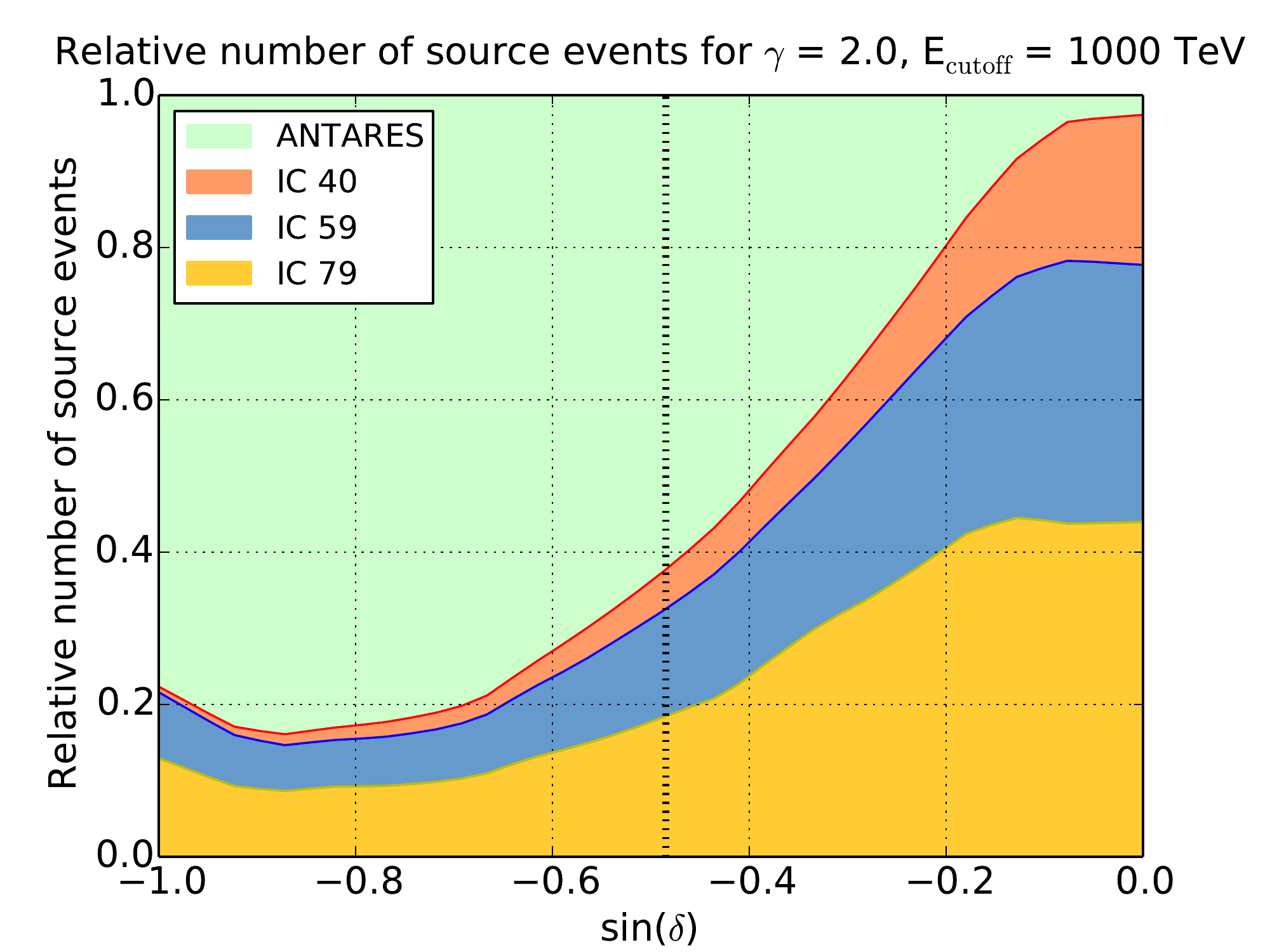}{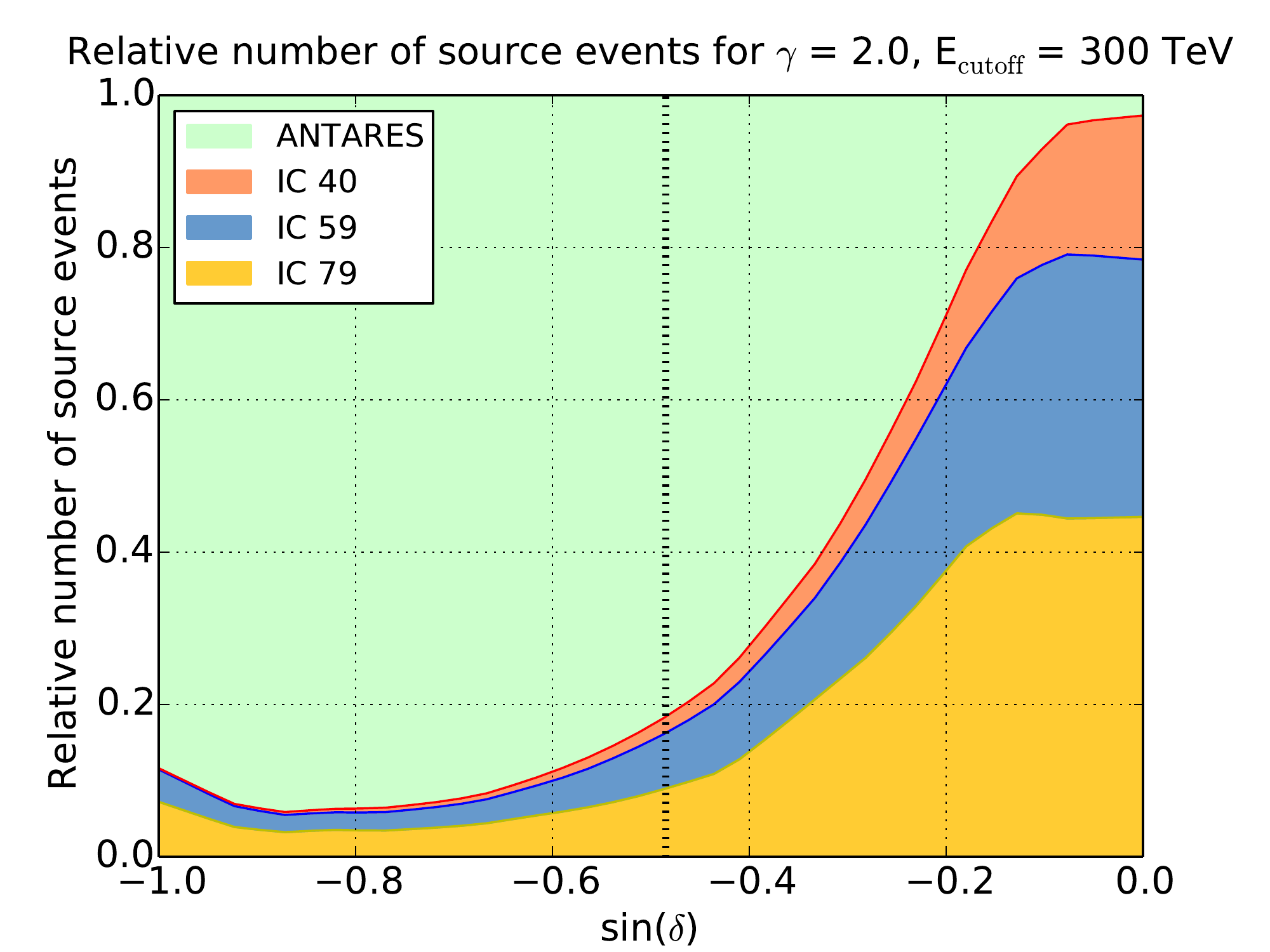}
    \plottwo{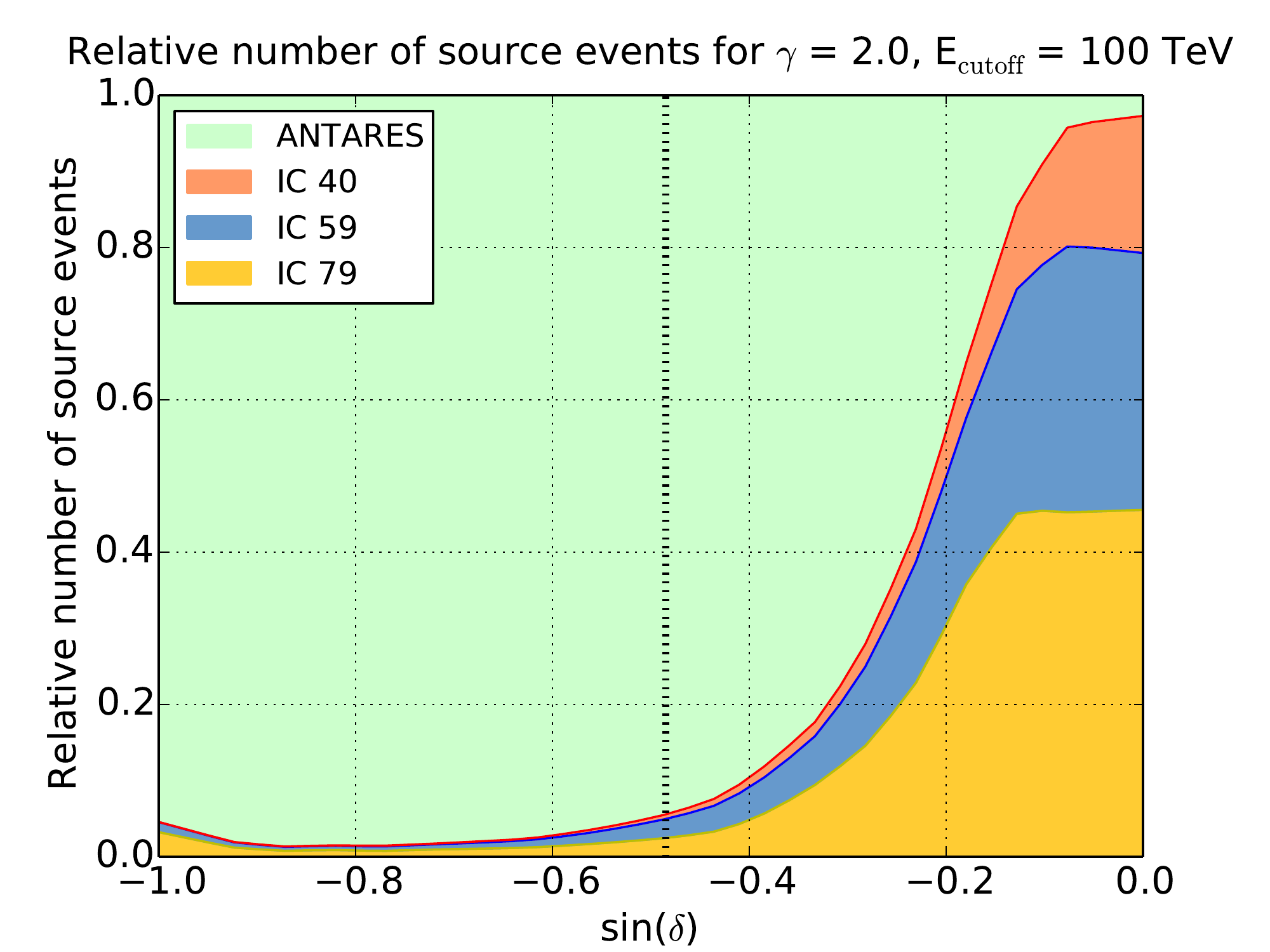}{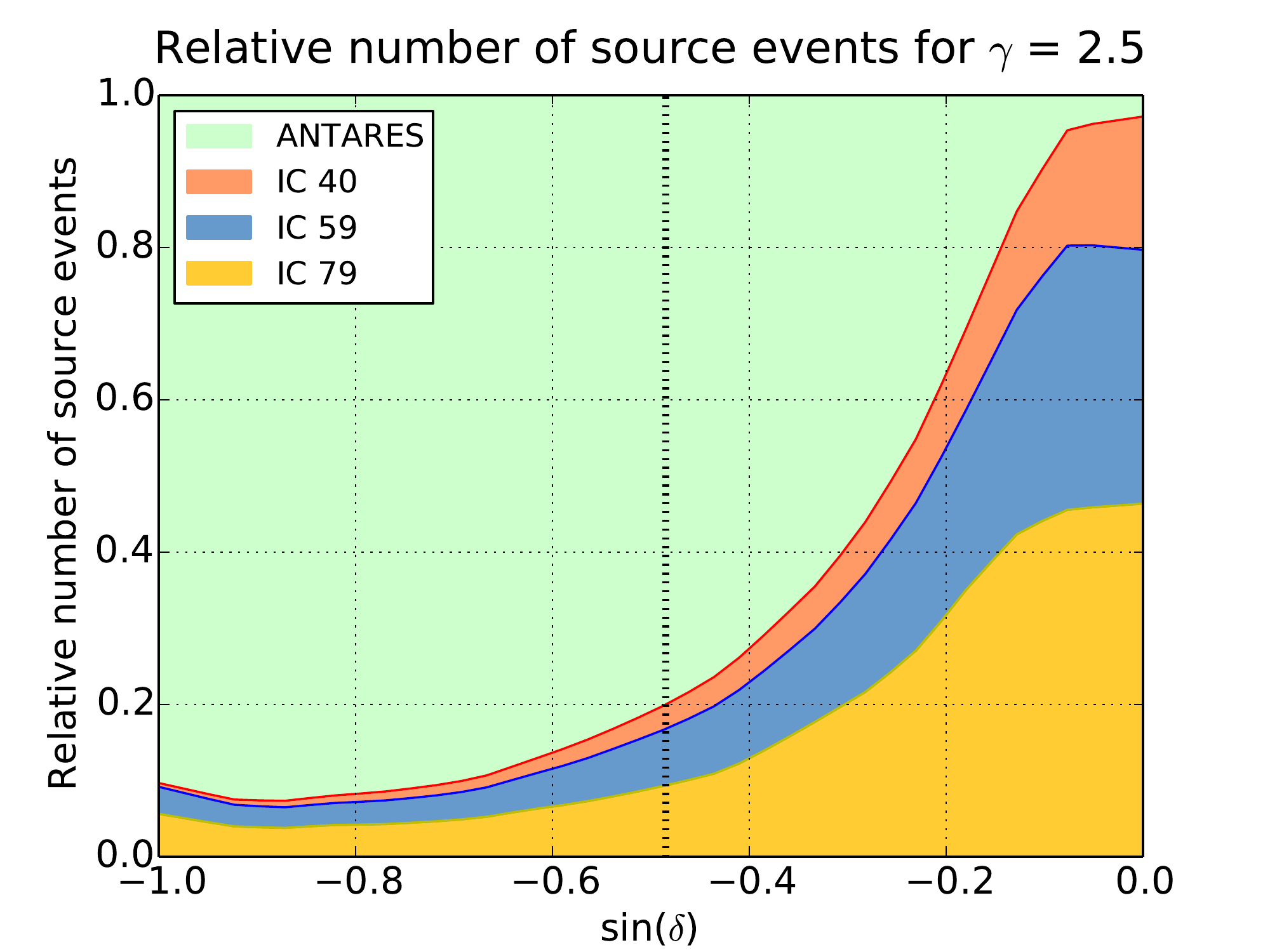}
	\caption{Relative fraction of signal events of each sample as a function of the source declination for different energy spectra: $E^{-2}$ with energy cutoff $E_{\rm cutoff}$ of 1 PeV (top-left), 300 TeV (top-right), 100 TeV (bottom-left); and $E^{-2.5}$ spectrum (bottom-right). The orange, blue and yellow shaded areas correspond to the IceCube 40, 59 and 79-string data samples, respectively, and the green shaded area indicates the ANTARES 2007-2012 sample. The vertical dashed line corresponds to the declination of the Galactic Center.}
	\label{RelContCases}
\end{figure}

Other source assumptions are also considered in this analysis. The relative fraction of source events is calculated  for an unbroken $E^{-2.5}$ power-law spectrum, as suggested in recent IceCube diffuse-flux searches \citep{IceCube-Diffuse}, and for an E$^{-2}$ spectrum with  exponential square-root cut-offs
 ($\frac{d\Phi}{dE} \propto E^{-2} \exp \left[ { - \sqrt{ \frac{E}{E_{\rm{cut-off}} } } } \right] $)
 of 100 TeV, 300 TeV and 1 PeV, since a square-root dependence may be expected from Galactic sources \citep{KAPPES}. Figure \ref{RelContCases} shows the relative fraction of source events for these cases. Compared with an unbroken $E^{-2}$ spectrum, the contribution of high energy neutrinos in all of these cases is lower, and therefore the relative contribution of the ANTARES sample increases.

\section{Search method}\label{method}

An unbinned maximum likelihood ratio estimation has been performed to search for excesses of events that would indicate cosmic neutrinos coming from a common source. In order to estimate the significance of a cluster of events, this likelihood takes into account the energy and directional information of each event. Due to the different detector response, the  data sample which an event belongs to is also taken into account. The likelihood, as a function of the total number of fitted signal events, $n_s$, can be expressed as:

\begin{equation}\label{eq:likelihood}
	L(n_s) = \prod_{j=1}^4 \prod_{i=1}^{N^j} \left[ \frac{n^j_s}{N^j}S^j_i + \left(1-\frac{n^j_s}{N^j} \right) B^j_i \right]
\end{equation}

\par \noindent where $j$ marks one of the four data samples, $i$ indicates an event belonging to the $j$-th sample, S$_i^j$ is the value of the signal probability distribution function (PDF) for the $i$-th event, B$^j_i$ indicates the value of the background PDF, $N^j$ is the total number of events in the $j$-th sample, and $n^j_s$ is the number of signal events fitted for in the $j-$th sample. Since a given evaluation of the likelihood refers to a single source hypothesis at a fixed sky location, the number of signal events $n^j_s$ that is fitted for in each sample is related to the total number of signal events $n_s$ by the relative contribution of each sample, $n^j_s = n_s \cdot C^{j}(\delta, \frac{d{\Phi}}{dE})$.

The signal and background PDFs for the IceCube and ANTARES samples have slightly different definitions. The signal PDF for ANTARES is defined as

\begin{equation}
	S^{ANT} = \frac{1}{2\pi\sigma^2} \exp\left(-\frac{\Delta \Psi(\vec{x}_s)^2}{2\sigma^2}\right) P^{ANT}_{s}(\mathcal{N}^{hits}, \sigma) ,
\end{equation}

\par \noindent where $\vec{x_s}$ = ($\alpha_s$, $\delta_s$) indicates the source direction in equatorial coordinates, $\Delta \Psi(\vec{x}_s)$ is the angular distance of a given event to the source and P$_s^{ANT}(\mathcal{N}^{hits}, \sigma)$ is the probability for a signal event to be reconstructed with an angular error estimate of $\sigma$ and a number of hits $\mathcal{N}^{hits}$. The number of hits is a proxy for the energy of the event. In this sense, an event with a higher number of hits (higher deposited energy) would be less likely to be produced by the expected background.

The definition of the signal PDFs for the IceCube samples is similar,

\begin{equation}
	S^{IC} = \frac{1}{2\pi\sigma^2} \exp\left(-\frac{\Delta \Psi(\vec{x}_s)^2}{2\sigma^2}\right) P^{IC}_{s}(\mathcal{E}, \sigma|\delta)
\end{equation}

\par \noindent where the main difference lies in the use of the reconstructed energy, $\mathcal{E}$, and the declination dependence of the probability for a signal event to be reconstructed with a given $\sigma$ and $\mathcal{E}$. The declination dependence is needed mainly because of the event selection cut on reconstructed energy, which is designed to reduce the atmospheric muon background. 

Background events are expected to be distributed uniformly in right ascension. The background PDFs are in fact obtained from the experimental data itself. The definition of the PDFs are:

\begin{equation}
	B^{ANT} = \frac{B^{ANT}(\delta)}{2\pi} P^{ANT}_{b}(\mathcal{N}^{hits}, \sigma) , 
\end{equation}

\begin{equation}
	B^{IC} = \frac{B^{IC}(\delta)}{2\pi} P^{IC}_{b}(\mathcal{E}, \sigma|\delta) ,
\end{equation}

\par \noindent where B$^j$($\delta$) is the per-solid-angle rate of observed events as a function of the declination in the corresponding sample. $P^{ANT}_{b}(\mathcal{N}^{hits}, \sigma)$ and $P^{IC}_{b}(\mathcal{E}, \sigma|\delta)$ characterize the distributions for background event properties, in analogy with the definitions of $P^{ANT}_{s}$ and $P^{IC}_{s}$ for signal events given above.

The test statistic, TS, is determined from the likelihood (Eq.~\ref{eq:likelihood}) as TS = $\log L(\hat{n}_s) - \log L(n_s=0)$, where $\hat{n}_s$ is the value that maximizes the likelihood. The larger the TS, the lower the probability (p-value) of the observation to be produced by the expected background. Simulations are performed to obtain the distributions of the TS. The significance (specifically, the p-value) of an observation is determined by the fraction of TS values which are larger than the observed TS.

The TS is calculated as a preliminary step to obtain the post-trial p-values of a search. TS distributions for the fixed-source, background-only hypothesis have been calculated in steps of 1$^\circ$ in declination from pseudo-data sets of randomized data. Because these distributions vary with declination, the preliminary TS is turned into a ``pre-trial p-value'' by comparing the TS obtained at the source location from the data to the background TS distribution for the corresponding declination.  The post-trial significance is then estimated with pseudo-data sets and according to the type of search, as explained together with the results in Section \ref{results}.

Two different searches for point-like neutrino sources have been performed. In the candidate list search, a possible excess of neutrino events is looked for at the location of 40 pre-selected neutrino source candidates. Since the location of these sources is fixed (at known locations with an uncertainty below the angular resolution of all samples) only the number of signal events $n_s$ is a free parameter in the likelihood maximisation. These candidates correspond to all sources in the Southern sky considered in the previous candidate-source list searches performed in the ANTARES and IceCube point-source analyses \citep{PS-IceCube-79, PS-ANTARES}.

The second search is a ``full sky'' search, looking for a significant point-like excess anywhere in the Southern sky. For this purpose, the likelihood is evaluated in steps of 1$^\circ \times $~1$^\circ$ over the whole scanned region. In this case, the source position is an additional free parameter of the likelihood to fit the best position within the 1$^\circ \times$~1$^\circ$ boundaries. 

Both the full Southern-sky and candidate-list searches have been performed using an $E^{-2}$
source spectrum in the signal PDFs. The main virtue of the energy term in the PDFs is to add power to distinguish signal neutrinos from the softer spectra of atmospheric neutrinos and atmospheric muons. Limits for the sources in the candidate list have also been calculated for the source spectra mentioned in Section \ref{Sec:Relative}.

\section{Results}\label{results}

Results from the full Southern sky and candidate list searches are detailed below.

\subsection{Full Southern-sky search}

\begin{figure}[!h]
	\centering
	\includegraphics[width=.75\textwidth]{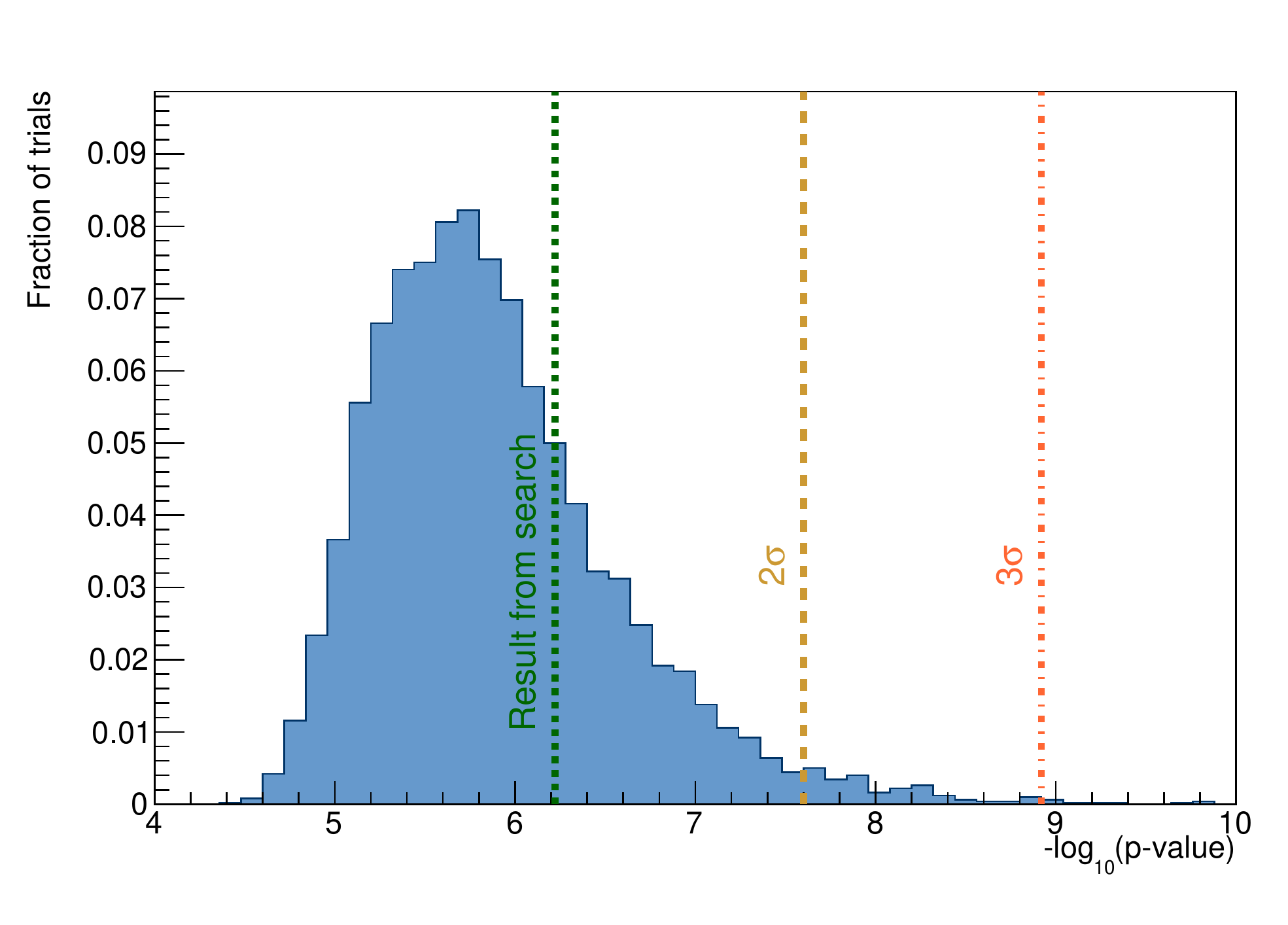}
	\caption{Distribution of the smallest p-value in the Southern sky obtained from scans of pseudo-data sets. Green line: pre-trial p-value for the most significant source location found. Yellow and red lines: pre-trial p-values for the 2$\sigma$ and 3$\sigma$ significance thresholds using the one-sided sigma convention.}
	\label{fullskysearch}
\end{figure}

\begin{figure}[!h]
	\centering
	\plotone{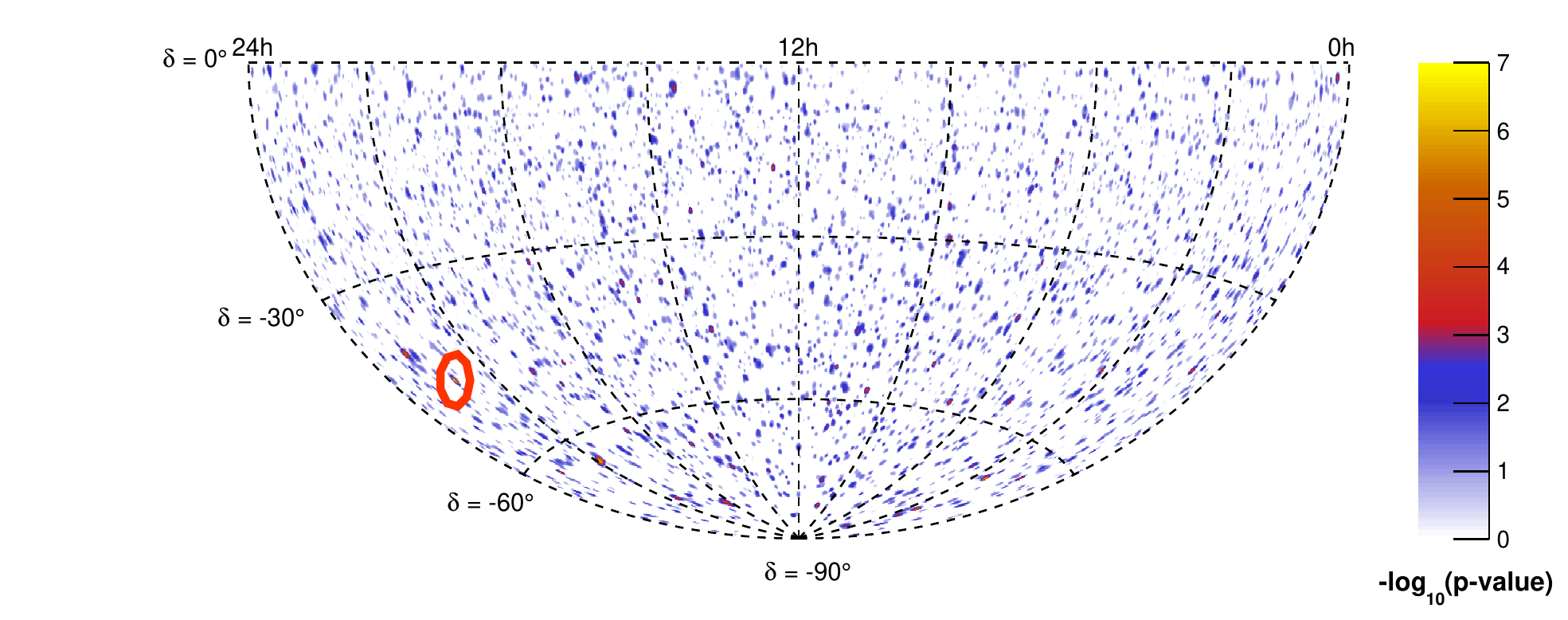}
	\caption{Skymap of pre-trial p-values for the combined ANTARES 2007-2012 and IceCube 40, 59, 79 point-source analyses. The red circle indicates the location of the most significant cluster ($0.7\sigma$ post-trial significance in the one-sided sigma convention), discussed in the text.}
	\label{pretrialmap}
\end{figure}

No significant event clusters are found over the expected background. The most significant cluster is located at equatorial coordinates $\alpha$ = 332.8$^\circ$, $\delta$=--46.1$^\circ$, with best-fit $n_s=7.9$ and pre-trial p-value of $6.0\times10^{-7}$.  Figure \ref{fullskysearch} shows this pre-trial p-value compared to the distribution of smallest p-values found when performing the same analysis on many pseudo-data sets (constructed by randomizing the right ascension coordinates of the real data).  It is found that 24\% of pseudo-data sets have a smaller p-value somewhere in the sky than is found in the real data;  the post-trial significance is thus 24\% (0.7$\sigma$ in the one-sided sigma convention).  The direction of this cluster is consistent with, but also less significant than the second most significant cluster in the previous ANTARES point-source analysis.  Figure \ref{pretrialmap} shows the position of this cluster and the pre-trial p-values for all directions in the Southern sky (a smaller step of 0.2$^\circ \times$0.2$^\circ$ is used to plot this map).

\subsection{Candidate list search}

\begin{figure}[!h]
	\centering
	\includegraphics[width=.75\textwidth]{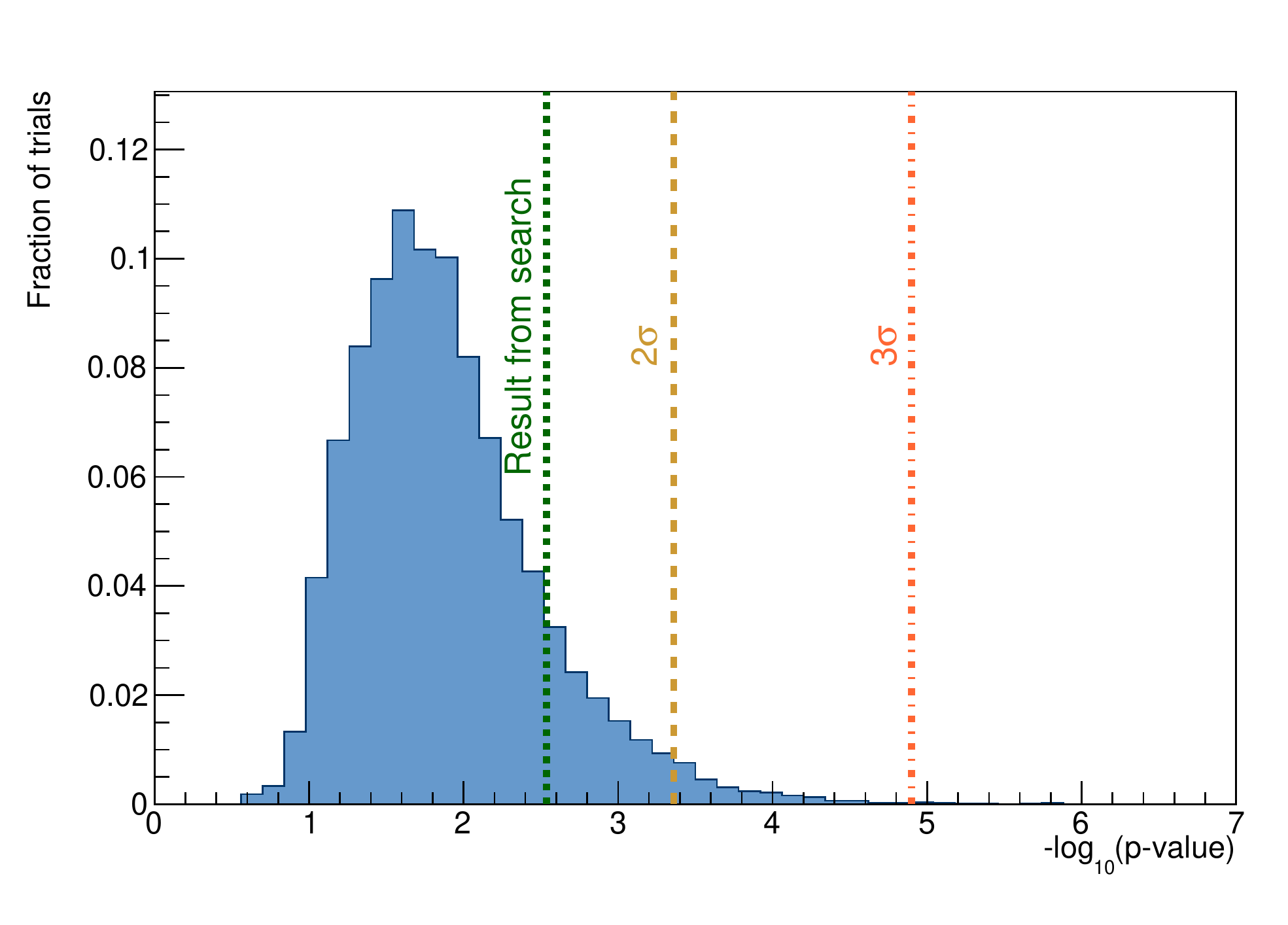}
	\caption{Distribution of the smallest p-value found in each candidate-list analysis of a pseudo-data set. Green line: pre-trial p-value for the most significant object found in the real data. Yellow and red lines: pre-trial p-values needed for the 2 and 3$\sigma$ post-trial significance thresholds in the one-sigma convention.}
	\label{CLsearch}
\end{figure}

The results of the candidate source list search are presented in Table\,\ref{tab:CL}.  No statistically significant excess is found.  The most significant excess for any object on the list corresponds to HESS\,J1741-302 with a pre-trial p-value of 0.003.
 To account for trial factors, the search is performed on the same list of sources using pseudo data-sets, and the distribution of smallest p-values for these searches is shown in Figure\,\ref{CLsearch}.  We find that 11\% of randomized data sets have a smaller p-value for any source than that found for the real data; the post-trial significance of the source list search is thus 11\% ($1.2\sigma$ in the one-sided sigma convention).  

\begin{figure}[!htp]
	\centering
	\includegraphics[width=.8\textwidth]{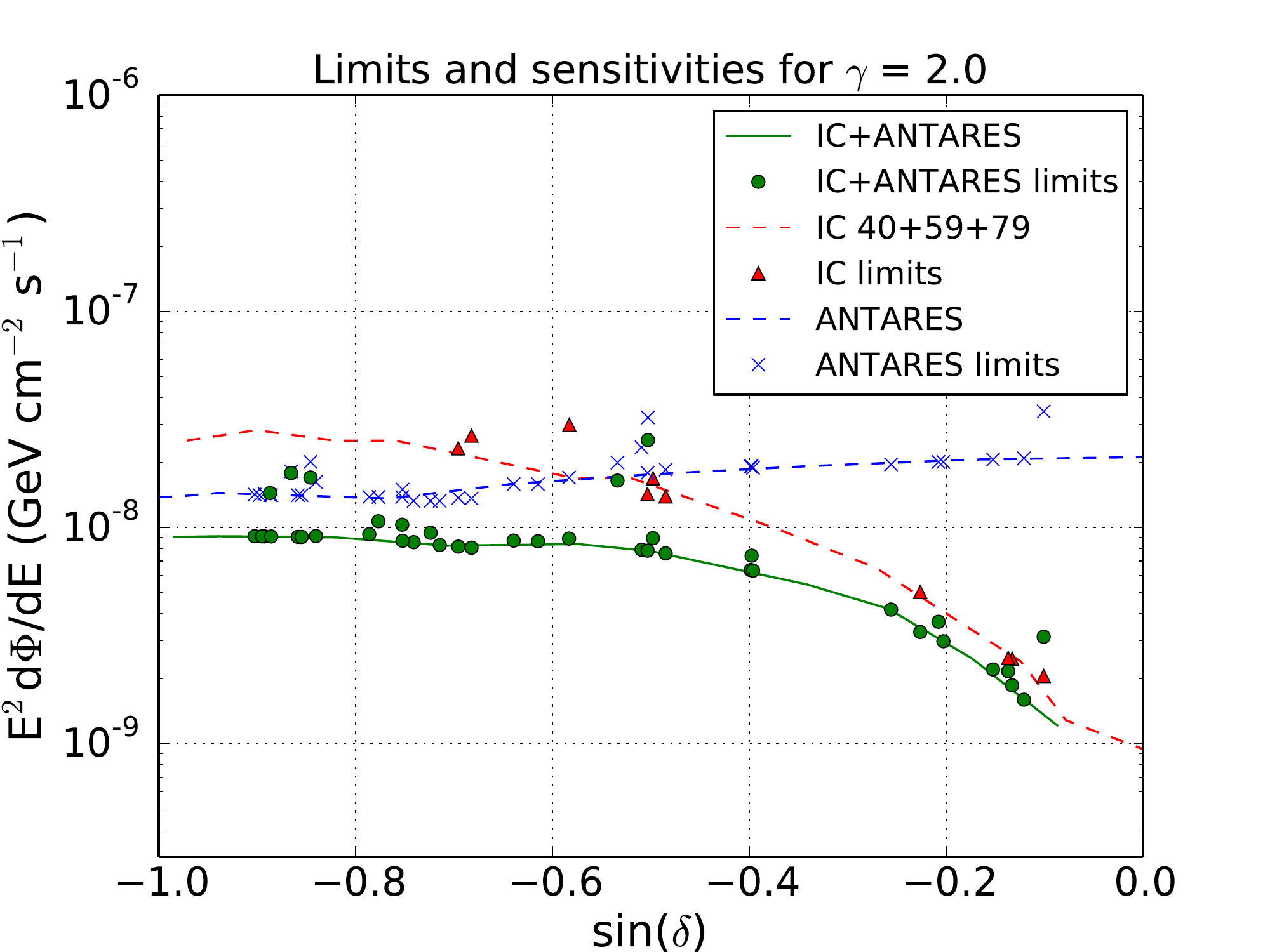}
	\caption{90\% CL sensitivities and limits (Neyman method) for the neutrino emission from point sources as a function of source declination in the sky, for an assumed $E^{-2}$ energy spectrum of the source.  Green points indicate the actual limits on the candidate sources. The green line indicates the sensitivity of the combined search. Curves/points respectively indicate the published sensitivities/limits for the IceCube (blue) and ANTARES (red) analyses, respectively. As reference, the declination of the Galactic Center is approximately at sin($\delta = -29^\circ$) $\approx$ -0.48. }
	\label{sensX20}
\end{figure}

\begin{figure}[!tph]
	\centering
	\plottwo{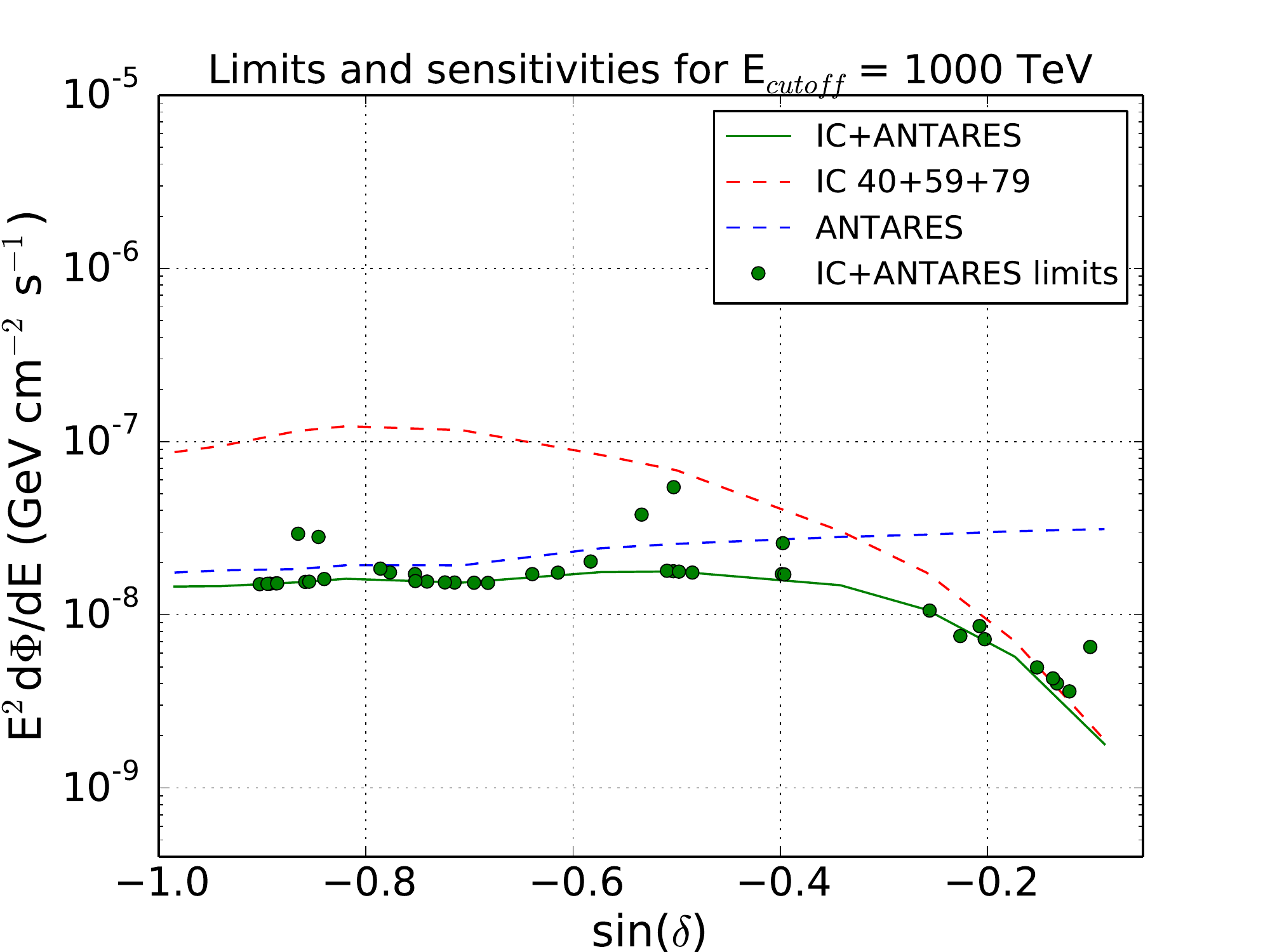}{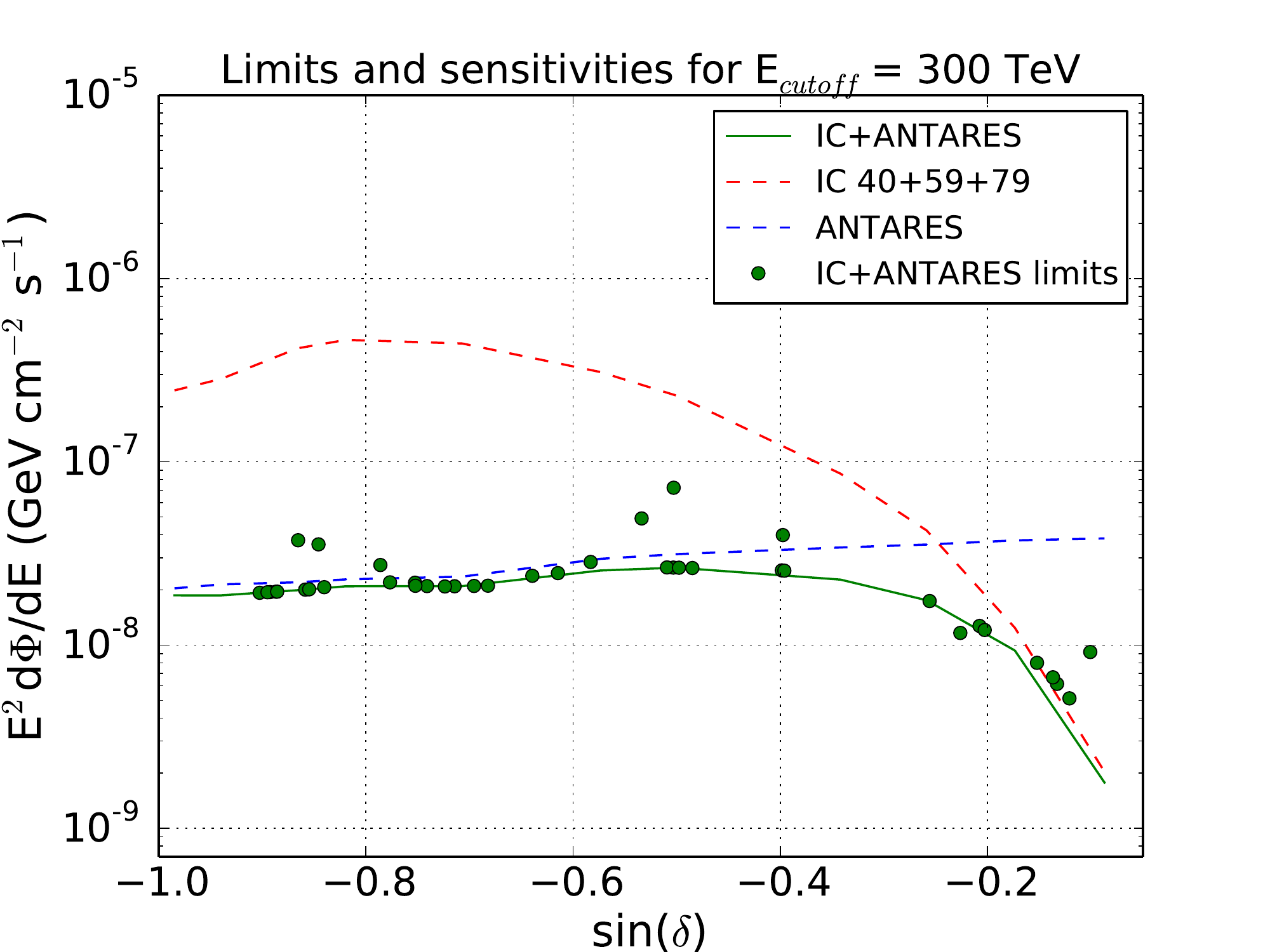}
    \plottwo{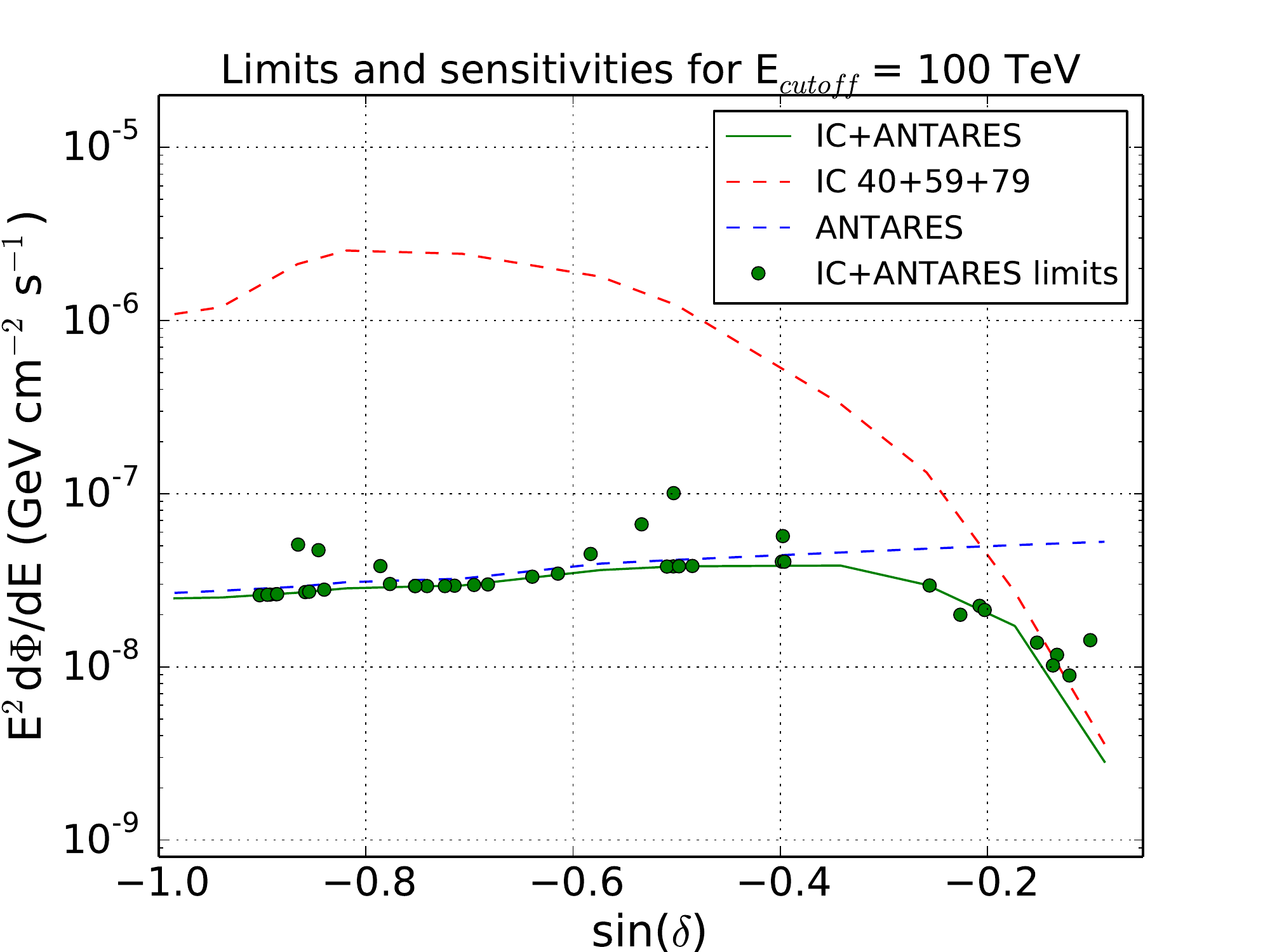}{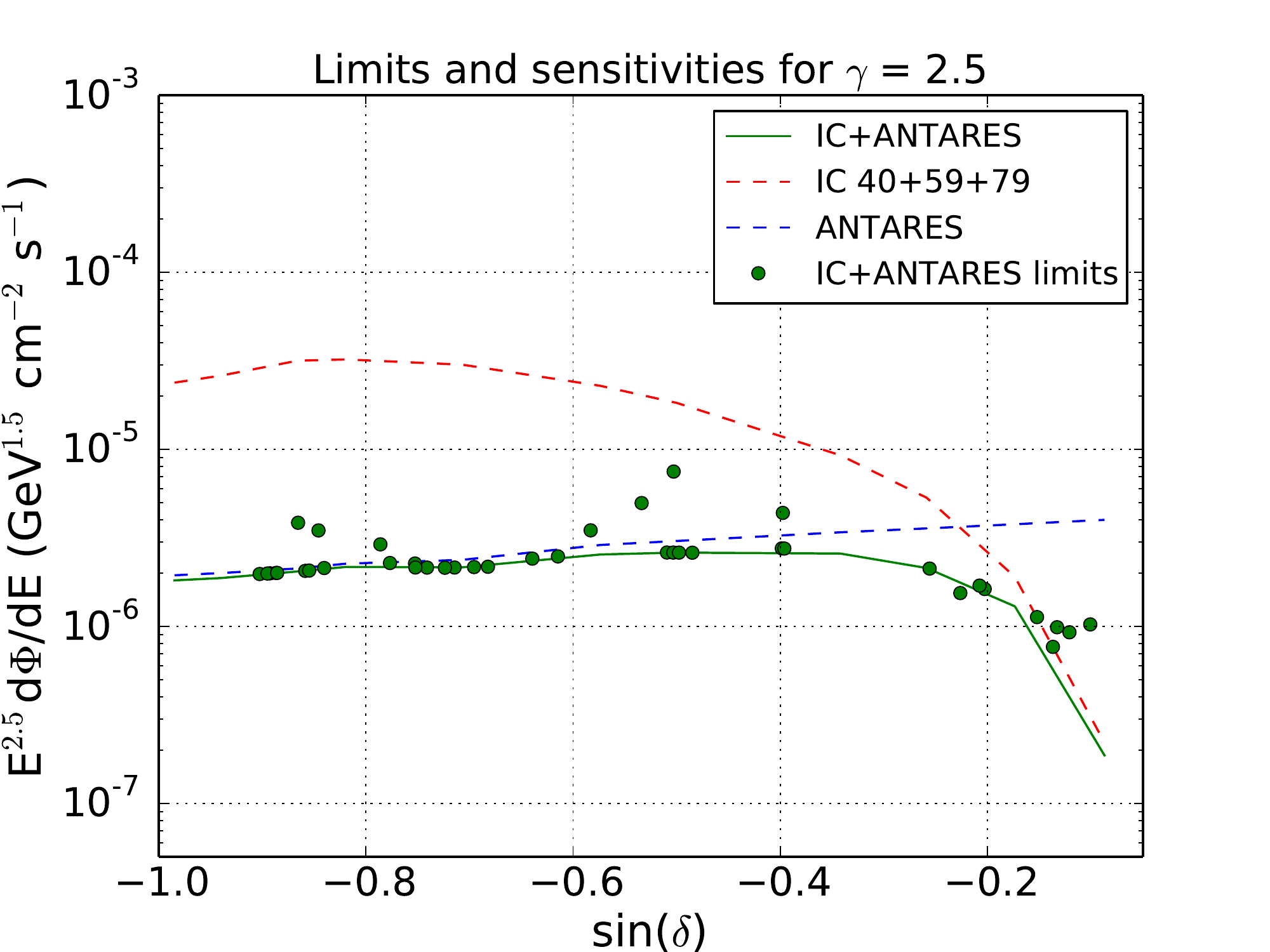}
	\caption{Point source sensitivities and limits as in Fig.\,\ref{sensX20}, for other energy spectra: $E^{-2}$ with a square-root exponential cut-off at $E = 1$\,PeV (top left), $E = 300$\,TeV (top right), $E = 100$\,TeV (bottom left) and $E^{-2.5}$ unbroken power-law (bottom right). Green points indicate the actual limits on the candidate sources. The green line indicates the sensitivity for the combined search. Red and blue curves indicate the sensitivities for the individual IceCube and ANTARES analyses, respectively. As reference, the declination of the Galactic Center is approximately at sin($\delta = -29^\circ$) $\approx$ -0.48. }
	\label{sensCases}
\end{figure}

Table \ref{tab:CL} provides the pre-trial p-values, best-fit signal events $n_s$ and flux upper limits (under different assumptions of the energy spectrum) for all the candidate source objects.  Figure \ref{sensX20} shows the Neyman sensitivities and limits for this search (assuming an E$^{-2}$ spectrum) in comparison with the previously published ANTARES and IceCube analyses of the same data.  The point-source sensitivity in a substantial region of the sky, centered approximately at the declination of the Galactic center ($\delta=-29^{\circ}$), can be seen to have improved by up to a factor of two. A maximum gain of at most $\sqrt{2}$ would be expected in a background-dominated sample; however, the low number of effective background events (with reconstructed energy and direction mimicking an astrophysical neutrino) is very low, so that gains of more than $\sqrt{2}$ are possible. Similar gains in other regions of the sky can be seen for different energy spectra in Figure \ref{sensCases}.

\begin{deluxetable}{lccccccccc}
\tabletypesize{\scriptsize}
\tablewidth{0pt}
\tablecaption{\footnotesize Fitted number of source events, $n_s$, pre-trial p-values, $p$, and 90\% C.L. flux limits, $\Phi^{90CL}_{\nu}$ for the different source spectra for the 40 candidate sources. Units for the flux limits for the E$^{-2.5}$ spectra, $\phi^{90CL}_{E^{-2.5}}$, are given in GeV$^{1.5}$cm$^{-2}$s$^{-1}$, whereas the rest are in GeV cm$^{-2}$s$^{-1}$. The sources are sorted by their declination. Dashes (-) in the fitted number of source events and pre-trial p-values indicate sources with $n_s \leq$ 0.001.  } 
\tablehead{
\colhead{Name}           			  & \colhead{$\delta$ ($^\circ$)}      &
\colhead{$\alpha$ ($^\circ$)}          & \colhead{$n_s$ }  &
\colhead{$p$}          & \colhead{$\phi^{90CL}_{ E^{-2}}$ } & 
\colhead{$\phi^{90\%CL}_{E_{c} = 1PeV}$  } & \colhead{$\phi^{90CL}_{E_{c} = 300 TeV}$ } &
\colhead{$\phi^{90CL}_{E_{c} = 100 TeV}$} & \colhead{$\phi^{90CL}_{E^{-2.5}}$  } }
\startdata
3C279   & -5.8  & -166.0        & 1.1   & 0.05  & 3.1E-09       & 1.0E-06       & 6.5E-09       & 9.2E-09       & 6.7E-08\\
HESSJ1837-069   & -7.0  & -80.6         & -     & -     & 1.6E-09       & 9.3E-07       & 1.5E-08       & 2.0E-08       & 2.6E-08\\
QSO2022-077     & -7.6  & -53.6         & -     & -     & 1.9E-09       & 1.7E-06       & 1.7E-08       & 2.5E-08       & 3.5E-08\\
PKS1406-076     & -7.9  & -147.8        & -     & -     & 2.2E-09       & 7.7E-07       & 4.3E-09       & 6.7E-09       & 1.0E-08\\
HESSJ1834-087   & -8.8  & -81.3         & -     & -     & 2.2E-09       & 1.1E-06       & 1.6E-08       & 2.1E-08       & 2.9E-08\\
PKS0727-11      & -11.7         & 112.6         & -     & -     & 3.0E-09       & 9.9E-07       & 1.5E-08       & 5.1E-09       & 8.9E-09\\
1ES0347-121     & -12.0         & 57.4  & -     & -     & 3.7E-09       & 2.5E-06       & 1.8E-08       & 2.6E-08       & 3.8E-08\\
QSO1730-130     & -13.1         & -96.7         & -     & -     & 3.3E-09       & 2.3E-06       & 1.8E-08       & 2.6E-08       & 3.8E-08\\
LS5039  & -14.8         & -83.4         & -     & -     & 4.2E-09       & 2.1E-06       & 1.1E-08       & 1.7E-08       & 2.9E-08\\
W28     & -23.3         & -89.6         & -     & -     & 6.3E-09       & 2.8E-06       & 1.7E-08       & 2.5E-08       & 4.0E-08\\
PKS0454-234     & -23.4         & 74.3  & -     & -     & 7.4E-09       & 2.2E-06       & 2.6E-08       & 2.7E-08       & 3.8E-08\\
1ES1101-232     & -23.5         & 165.9         & -     & -     & 6.4E-09       & 2.6E-06       & 7.2E-09       & 1.2E-08       & 2.1E-08\\
Galactic Center  & -29.0         & -93.6         & -     & -     & 7.6E-09       & 2.6E-06       & 1.8E-08       & 2.6E-08       & 3.8E-08\\
PKS1622-297     & -29.9         & -113.5        & -     & -     & 8.9E-09       & 2.6E-06       & 1.7E-08       & 2.2E-08       & 2.9E-08\\
HESSJ1741-302   & -30.2         & -94.8         & 1.6   & 0.003         & 2.5E-08       & 7.5E-06       & 5.5E-08       & 7.2E-08       & 1.0E-07\\
PKS2155.304     & -30.2         & -30.3         & -     & -     & 7.8E-09       & 2.6E-06       & 2.0E-08       & 2.8E-08       & 4.5E-08\\
H2356-309       & -30.6         & -0.2  & -     & -     & 7.9E-09       & 1.5E-06       & 1.5E-08       & 2.1E-08       & 3.0E-08\\
PKS0548-322     & -32.3         & 87.7  & 0.9   & 0.07  & 1.6E-08       & 5.0E-06       & 3.8E-08       & 4.9E-08       & 1.4E-08\\
PKS1454-354     & -35.7         & -135.6        & -     & -     & 8.9E-09       & 3.5E-06       & 8.6E-09       & 2.1E-08       & 3.0E-08\\
PKS0426-380     & -37.9         & 67.2  & -     & -     & 8.6E-09       & 2.8E-06       & 7.5E-09       & 1.2E-08       & 2.0E-08\\
RXJ1713.7-3946  & -39.8         & -101.8        & -     & -     & 8.7E-09       & 2.0E-06       & 1.5E-08       & 2.0E-08       & 2.6E-08\\
CenA    & -43.0         & -158.6        & -     & -     & 8.1E-09       & 2.2E-06       & 4.0E-09       & 6.1E-09       & 1.2E-08\\
PKS0537-441     & -44.1         & 84.7  & -     & -     & 8.2E-09       & 1.6E-06       & 1.8E-08       & 2.6E-08       & 4.1E-08\\
VelaX   & -45.6         & 128.8         & -     & -     & 8.3E-09       & 2.2E-06       & 1.5E-08       & 2.1E-08       & 2.9E-08\\
RXJ0852.0-4622  & -46.4         & 133.0         & -     & -     & 9.5E-09       & 2.1E-06       & 1.5E-08       & 2.1E-08       & 2.9E-08\\
HESSJ1632-478   & -47.8         & -112.0        & -     & -     & 8.6E-09       & 2.1E-06       & 1.6E-08       & 2.1E-08       & 2.9E-08\\
PKS2005-489     & -48.8         & -57.6         & -     & -     & 1.0E-08       & 2.9E-06       & 1.7E-08       & 1.3E-08       & 2.2E-08\\
GX339-4         & -48.8         & -104.3        & -     & -     & 8.7E-09       & 2.2E-06       & 1.6E-08       & 2.1E-08       & 2.8E-08\\
HESSJ1616-508   & -51.0         & -116.0        & -     & -     & 1.1E-08       & 2.3E-06       & 1.8E-08       & 2.2E-08       & 3.0E-08\\
HESSJ1614-518   & -51.8         & -116.4        & -     & -     & 9.3E-09       & 2.1E-06       & 1.6E-08       & 2.0E-08       & 2.7E-08\\
CirX-1  & -57.2         & -129.8        & -     & -     & 9.1E-09       & 2.1E-06       & 1.8E-08       & 2.7E-08       & 3.8E-08\\
HESSJ1023-575   & -57.8         & 155.8         & 0.8   & 0.08  & 1.7E-08       & 3.5E-06       & 2.8E-08       & 3.5E-08       & 4.7E-08\\
HESSJ1503-582   & -58.7         & -133.6        & -     & -     & 9.1E-09       & 2.0E-06       & 1.5E-08       & 1.9E-08       & 2.6E-08\\
MSH15-52        & -59.2         & -131.5        & -     & -     & 9.1E-09       & 2.1E-06       & 1.5E-08       & 2.0E-08       & 2.7E-08\\
ESO139-G12      & -59.9         & -95.6         & 0.8   & 0.07  & 1.8E-08       & 3.9E-06       & 2.9E-08       & 3.7E-08       & 5.1E-08\\
HESSJ1507-622   & -62.3         & -133.3        & -     & -     & 9.1E-09       & 2.0E-06       & 5.0E-09       & 8.0E-09       & 1.4E-08\\
RCW86   & -62.5         & -139.3        & 0.2   & 0.11  & 1.4E-08       & 4.4E-06       & 3.6E-09       & 4.0E-08       & 5.7E-08\\
HESSJ1303-631   & -63.2         & -164.2        & -     & -     & 9.1E-09       & 2.0E-06       & 1.5E-08       & 1.9E-08       & 2.6E-08\\
PSRB1259-63     & -63.5         & -164.3        & -     & -     & 9.1E-09       & 2.4E-06       & 1.7E-08       & 2.4E-08       & 3.3E-08\\
HESSJ1356-645   & -64.5         & -151.0        & -     & -     & 9.1E-09       & 2.0E-06       & 1.5E-08       & 1.9E-08       & 2.6E-08\\
\enddata
\label{tab:CL}
\end{deluxetable}

\section{Conclusion}\label{conclusions}

We have presented the first combined point-source analysis of data from the ANTARES and IceCube detectors.  Their different characteristics, in particular IceCube's larger size  and ANTARES' location in the Northern hemisphere, complement each other for Southern sky searches. We have calculated the sensitivity to point sources and, with respect to an analysis of either data set alone, found that up to a factor of two improvement is achieved in different regions of the Southern sky, depending on the energy spectrum of the source.  Two joint analyses of the data sets have been performed: a search over the whole Southern sky for a point-like excess of neutrino events, and a targeted analysis of 40 pre-selected candidate source objects.  The largest excess in the Southern sky search has a post-trial probability of 24\% (significance of 0.7$\sigma$), located at $\alpha$ = 332.8$^\circ$, $\delta$=--46.1$^\circ$ in equatorial coordinates. In the source list search, the candidate with the highest significance corresponds to HESS\,J1741-302, with a post-trial probability of 11\% (significance of 1.2$\sigma$). Both of the results are compatible with the background-only hypothesis. Flux upper limits for each of the source candidates have been calculated for $E^{-2}$ and $E^{-2.5}$ power-law energy spectra, as well as for $E^{-2}$ spectra with cut-offs at energies of 1\,PeV, 300\,TeV, and 100\,TeV.  Because of their complementary nature, with IceCube providing more sensitivity at higher energies and ANTARES at lower energies, a joint analysis of future data sets will continue to provide the best point-source sensitivity in critical overlap regions in the Southern sky, where neutrino emission from Galactic sources in particular may be found.

\acknowledgments


The authors of the ANTARES collaboration acknowledge the financial support of the funding agencies:
Centre National de la Recherche Scientifique (CNRS), Commissariat \`a
l'\'ener\-gie atomique et aux \'energies alternatives (CEA),
Commission Europ\'eenne (FEDER fund and Marie Curie Program), R\'egion
\^Ile-de-France (DIM-ACAV) R\'egion Alsace (contrat CPER), R\'egion
Provence-Alpes-C\^ote d'Azur, D\'e\-par\-tement du Var and Ville de La
Seyne-sur-Mer, France; Bundesministerium f\"ur Bildung und Forschung
(BMBF), Germany; Istituto Nazionale di Fisica Nucleare (INFN), Italy;
Stichting voor Fundamenteel Onderzoek der Materie (FOM), Nederlandse
organisatie voor Wetenschappelijk Onderzoek (NWO), the Netherlands;
Council of the President of the Russian Federation for young
scientists and leading scientific schools supporting grants, Russia;
National Authority for Scientific Research (ANCS), Romania; 
Mi\-nis\-te\-rio de Econom\'{\i}a y Competitividad (MINECO), Prometeo 
and Grisol\'{\i}a programs of Generalitat Valenciana and MultiDark, 
Spain; Agence de  l'Oriental and CNRST, Morocco. We also acknowledge 
the technical support of Ifremer, AIM and Foselev Marine for the sea 
operation and the CC-IN2P3 for the computing facilities.

The authors of the IceCube collaboration acknowledge the support from the following agencies:
U.S. National Science Foundation-Office of Polar Programs,
U.S. National Science Foundation-Physics Division,
University of Wisconsin Alumni Research Foundation,
the Grid Laboratory Of Wisconsin (GLOW) grid infrastructure at the University of Wisconsin - Madison, the Open Science Grid (OSG) grid infrastructure;
U.S. Department of Energy, and National Energy Research Scientific Computing Center,
the Louisiana Optical Network Initiative (LONI) grid computing resources;
Natural Sciences and Engineering Research Council of Canada,
WestGrid and Compute/Calcul Canada;
Swedish Research Council,
Swedish Polar Research Secretariat,
Swedish National Infrastructure for Computing (SNIC),
and Knut and Alice Wallenberg Foundation, Sweden;
German Ministry for Education and Research (BMBF),
Deutsche Forschungsgemeinschaft (DFG),
Helmholtz Alliance for Astroparticle Physics (HAP),
Research Department of Plasmas with Complex Interactions (Bochum), Germany;
Fund for Scientific Research (FNRS-FWO),
FWO Odysseus programme,
Flanders Institute to encourage scientific and technological research in industry (IWT),
Belgian Federal Science Policy Office (Belspo);
University of Oxford, United Kingdom;
Marsden Fund, New Zealand;
Australian Research Council;
Japan Society for Promotion of Science (JSPS);
the Swiss National Science Foundation (SNSF), Switzerland;
National Research Foundation of Korea (NRF);
Danish National Research Foundation, Denmark (DNRF).

{\it Facilities:} \facility{ANTARES, IceCube}.


\begin{thebibliography}{}

\bibitem[Aartsen et al. (2013a)]{IceCube1}  Aartsen~M.~G., Abbasi~R., Abdou~Y.  et al.  (IceCube Collaboration) 2013a,  Phys.\ Rev.\ Lett., 111, 021103.

\bibitem[Aartsen et al. (2013b)]{IceCube_HESE_3yr} Aartsen, M.~G., et al. (IceCube Collaboration) 2013b, Phys.\ Rev.\ Lett.,113, 101101.

\bibitem[Aartsen et al. (2013c)]{IceCube_HESE_2yr} Aartsen, M.~G., et al. (IceCube Collaboration) 2013c, Sci, 342, 1242856.

\bibitem[Aartsen et al. (2013d)]{PS-IceCube-79} Aartsen~M.~G., Abbasi~R., Abdou~Y.  et al.  (IceCube Collaboration)  2013d, ApJ,  779, 132. 

\bibitem[Aartsen et al. (2015)]{IceCube-Diffuse} Aartsen~M.~G., Ackermann~M.,  Adams~J. et al. (IceCube Collaboration), 2015, Phys. Rev. D, 91, 022001.

\bibitem[Abbasi et al. (2010)]{IceCube_DOM}Abbasi, R., Abdou, Y., Abu-Zayyad, T., et al. (IceCube Collaboration) 2010, NIMPA, 618, 139. 
 
\bibitem[Abbasi et al. (2011)]{PS-IceCube-40}Abbasi, R. et al. (IceCube Collaboration) 2011, ApJ, 732, 18.

\bibitem[Abbasi et al. (2012)]{DeepCore}Abbasi, R. et al. (IceCube Collaboration) 2012, ApP, 35, 10, 615.

\bibitem[Achterberg et al. (2006)]{icecube}Achterberg, A., Ackermann, M., Adams, J., et al. (IceCube Coll.) 2006, APh, 26, 155.

\bibitem[Adri\'{a}n-Mart\'{i}nez et al. (2012a)]{ANTARES-OSCI}   Adrian-Martinez~S., Samarai~I.Al, Albert~A. et al.  (ANTARES Collaboration) 2012a, Phys. Lett. B, 714, 224.

\bibitem[Adri\'{a}n-Mart\'{i}nez et al. (2012b)]{AartThesis}   Adrian-Martinez~S., Samarai~I.Al, Albert~A. et al.  (ANTARES Collaboration) 2012b, ApJ, 760, 53.

\bibitem[Adri\'{a}n-Mart\'{i}nez et al. (2013a)]{OtherAart}   Adrian-Martinez~S., Samarai~I.Al, Albert~A. et al.  (ANTARES Collaboration) 2013a, JCAP, 1303, 006.

\bibitem[Adri\'{a}n-Mart\'{i}nez et al. (2013b)]{ANTARES-MUAT}   Adrian-Martinez~S., Samarai~I.Al, Albert~A. et al.  (ANTARES Collaboration) 2013b, Eur. Phys. J. C, 73, 2606.

\bibitem[Adri\'{a}n-Mart\'{i}nez et al. (2014)]{PS-ANTARES} Adri\'{a}n-Mart\'{i}nez~S., Albert~A., Andr\'{e}~ M.,  et al.  (ANTARES Collaboration)   2014, ApJ, 786, L5.  

\bibitem[Ageron et al. (2011)]{AntDetect}  Ageron~M., Aguilar~J.~A., Samarai~I.~Al et al.  (ANTARES Collaboration) 2011,  Nucl.\ Instrum.\ Meth., A 656, 11.

\bibitem[Ahlers et al. (2015)]{GAL-IC} Ahlers~M., Bai~Y., Barger~V. \&  Lua~R., 2015, arXiv:1505.03156v1.

\bibitem[Ahrens et al. (2002)]{amanda} Ahrens~J.,  Andr\'{e}s~E., Bai X. et al. 2002, Phys. Rev. D, 66, 012005

\bibitem[Alvarez-Mu\~{n}iz \& Halzen (2002)]{SNRs1} Alvarez-Mu\~{n}iz~J. \& Halzen~F., 2002, ApJ,  576, L33.

\bibitem[Anchordoqui et al. (2014a)]{GAL-IC2} Anchordoqui~L.~A. et al., 2014a, Phys. Rev. D, 89, 083003.

\bibitem[Anchordoqui et al. (2014b)]{GAL-IC3} Anchordoqui~L.~A. et al., 2014b, Phys. Rev. D, 90, 123010.

\bibitem[Atoyan \& Dermer(2001)]{AGN1} Atoyan~A. \&  Dermer~C.D., 2001, Phys.Rev.Lett. 87, 221102.

\bibitem[Aynutdinov et al.(2008)]{baikal}  Aynutdinov~V., Avrorin A., Balkanov V. et al. 2008, Nucl.Instrum.Meth. A588, 99.

\bibitem[Bai et al. (2014)]{GAL-IC4} Bai~Y. et al.,Phys. Rev. D, 2015, 90, 063012. 

\bibitem[Bandford \& Ostriker (1978)]{Fermi1} Blandford,~R.D. \& Ostriker,~J.P, 1978, ApJ, 221, L29.

\bibitem[Becker (2008)]{Origin6} Becker~J.~K., 2008, Physics Reports, 458, 173.

\bibitem[Becker et al. (2005)]{AGN2} Becker~J.~K., Biermann~P.~L. \& Rhode~W., 2005, ApP, 23, 4, 355.

\bibitem[Becker et al. (2006)]{GRB1} Becker~J.~K., Stamatikos~M., Halzen~F. \& Rhode~W., 2006, ApP, 25, 2, 118.

\bibitem[Bednarek (2005)]{MicroQ1} Bednarek W., 2005, ApJ, 631, 1.

\bibitem[Cavasinni et al. (2006)]{SNRs2} Cavasinni~V., Grasso~D. \& Maccione~L., 2006, ApP,  26, 41.

\bibitem[Cholis \& Hooper (2013)]{EXTRAGAL1} Cholis~I. \& Hooper~D., 2013, JCAP, 06, 030.

\bibitem[Eichmann et al. (2012)]{AGN3} Eichmann~B., Schlickeiser~R. \& Rhode~W., 2012, AAS, 749, 2.

\bibitem[Fujita et al.(2015)]{SgrA} Fujita~Y., Kimura~S.~S. \& Murase~K., 2015, Phys. Rev. D, 92, 023001.


\bibitem[Gaisser et al. (1995)]{Origin4} Gaisser~T.~K., Halzen~F. \& Stanev~T., 1995, Phys.Rept., 258, 173

\bibitem[Gonz\'{a}lez-Garc\'{i}a et al (2014)]{SNRs3} Gonz\'{a}lez-Garc\'{i}a~M.~C., Halzen~F., Niro~V, 2014, ApP, 57, 39.

\bibitem[Guetta \& Amato (2003)]{SNRs4} Guetta~D. \& Amato~E., 2003, ApP, 19, 403.

\bibitem[Halzen \& Hooper (2002)]{Origin3} Halzen~F. \& Hooper~D., 2002, Rept. Prog. Phys. 65, 1025.

\bibitem[Halzen et al. (2006)]{SNRs5} Halzen~F., Kappes~A. \& Murchadha~A., 2008, Phys. Rev. D, 78, 063004.

\bibitem[H\"ummer et al. (2012)]{GRB2} H\"ummer~S., Baerwald~P. \& Winter~W., 2012, Phys. Rev. Lett., 108, 231101.

\bibitem[Kappes et al. (2007)]{KAPPES} Kappes~A., Hinton~J., Stegmann~C. \& Aharonian~F.~A., 2007, ApJ, 656, 870.

\bibitem[Kalashev et al. (2013)]{EXTRAGAL2} Kalashev~O.~E., Kusenko~A. \& Essey~W., 2013, Phys. Rev. lett., 111, 041103.

\bibitem[Kelner \& Aharonian (2008)]{Origin2} Kelner S.R. \& aharonian F.A., Phys. Rev. D, 2008, 78, 034013.

\bibitem[Kistler \& Beacom (2006)]{GC} Kistler~M.~D. \& Beacom~J.~F., 2006, Phys. Rev. D, 74, 063007.

\bibitem[Krymskii (1997)]{Fermi2} Krymskii~G.F., 1977, Soviet Physics Doklady, 22, 327.

\bibitem[Learned \& Mannheim (2000)]{Origin5} Learned~J.~G. \& Mannheim~K., 2000, Annu. Rev. Nucl. Part. S., 50, 679

\bibitem[Levinson \& Waxman (2001)]{MicroQ2} Levinson~A. \& Waxman E., 2001, Phys. Rev. Lett. 87, 171101.

\bibitem[Mandelartz \& Becker Tjus (2015)]{SNRs6} Mandelartz~M. \& Becker Tjus~J., 2015, ApP, 65, 80.

\bibitem[Mannheim (1995)]{AGN4} Mannheim~K., 1995, ApP, 3, 295.

\bibitem[M\'{e}sz\'{a}ros \& Waxman (2001)]{GRB3} M\'{e}sz\'{a}ros~P. \& Waxman~E., 2001, Phys. Rev. Lett., 87, 171102.

\bibitem[M\"{u}cke et al. (2003)]{AGN5} M\"{u}cke~A., Protheroe~R.~J., Engel~R., Rachen~J.~P. \& Stanev~T., 2003, ApP, 18, 6, 593. 

\bibitem[Murase (2006)]{GRB4} Murase ~K. et al, 2006, ApJ, 651, L5.

\bibitem[Murase \& Nagataki (2006)]{GRB4} Murase~K. \& Nagataki~S., 2006, Phys. Rev. Lett., 97, 051101.

\bibitem[Murase (2015)]{Origin} Murase, 2015, AIP Conf.Proc. 1666, 040006.

\bibitem[Nellen et al. (1993)]{AGN6} Nellen~L., Mannheim~K. \& Biermann ~P.~L., 1993, Phys. Rev. D., 47, 5720.

\bibitem[Neyman (1937)]{Neyman} Neyman~J., 1937, Phil. Trans. Royal Soc. London A, 236, 333.

\bibitem[Fox et al. (2013)]{GAL-IC5} Fox D. B.,  Kashiyama ~K. \& M\'eszar\'os~P., 2013, ApJ, 774, 74.

\bibitem[Padovani \& Resconi (2014)]{GAL-IC6} Padovani ~P. \& Resconi~E., 2014, MNRAS, 443, 474.

\bibitem[Rachen \& M\'{e}sz\'{a}ros (1998)]{AGN7} Rachen~J.~P. \& M\'{e}sz\'{a}ros ~P., 1998, Phys. Rev. D., 58, 123005. 

\bibitem[Roulet et al. (2013)]{EXTRAGAL3} Roulet~E., Sigi~G., van Vliet~A. \& Mollerach~S., 2013, JCAP, 01, 028.

\bibitem[Romero et al. (2003)]{MicroQ3} Romero~G.E., Torres~D.F., Kaufman Bernad\'{o} ~M.~M. \& Mirabel~I.~F., 2003, A\&A, 410, L1.

\bibitem[Razzaque et al. (2003)]{GRB5} Razzaque~S., M\'{e}sz\'{a}ros~P. \& Waxman~E., Phys. Rev. D, 68, 083001.

\bibitem[Razzaque (2013)]{GAL-IC7} Razzaque~S., 2013, Phys. Rev. D, 88, 081302.

\bibitem[Stecker et al. (1991)]{AGN8} Stecker~F.~W., Done~C., Salamon~M.~H. \& Sommers~P., 1991,  Phys. Rev. Lett., 66, 2697.

\bibitem[Stecker (2005)]{AGN9} Stecker~F.~W., 2005,  Phys. Rev. D, 72, 107301.

\bibitem[Stecker (2013)]{EXTRAGAL4} Stecker~F.~W., 2013, Phys. Rev. D, 88, 047301.

\bibitem[Torres et al. (2005)]{MicroQ4} Torres~D.F., Romero~G.E \& Mirabel~F., 2005, Chin. J. Astron. Astrophys., 5, 183.

\bibitem[Vissani et al. (2011)]{SNRs7} Vissani~F., Aharonian~F. \& Sahakyan~N., 2011, Astropart.Phys, 34, 778.

\bibitem[Waxman \& Bahcall(1997)]{GRB6} Waxman~E. \& Bahcall~J., 1997, Phys. Rev. Lett., 78, 2292.

\bibitem[Waxman \& Bahcall(1999)]{Waxman} Waxman~E. \& Bahcall~J., 1999, Phys.Rev.D, 59, 023002.

\bibitem[Waxman \& Bahcall(2000)]{GRB7} Waxman~E. \& Bahcall~J., 2000, ApJ, 541, 707.



\end{thebibliography}
\end{document}